\documentclass{aa}

\usepackage{graphicx}
\usepackage{txfonts}
\usepackage[]{hyperref}
\usepackage{xcolor}
\newcommand{\new}[1]{#1}

\begin{document}

\title{New ultracool dwarf candidates from multi-epoch WISE data}

\author{
    S.~Karpov\inst{1}
    \and
    O.~Malkov\inst{2}
    \and
    A.~Avdeeva\inst{2}
}

\institute{
Institute of Physics, Czech Acad. Sci., 182 21 Prague 8, Czech Republic
\email{karpov@fzu.cz}
\and
Institute of Astronomy, 48 Pyatnitskaya St., Moscow 119017, Russia
}

\date{Received September 15, 1996; accepted March 16, 1997}


\abstract
   {\new{Thirty years} after the discovery of brown dwarfs, the search for these objects continues, particularly in the vicinity of the Sun. Objects near the Sun are characterized by large proper motions, making them seen as fast-moving objects. While the Gaia DR3 catalogue is a comprehensive source of proper motions, it lacks the depth needed for discovering fainter objects. Modern multi-epoch surveys, with their greater depth, offer a new opportunity for systematic search for ultracool dwarfs.
   }
   {The study aims to systematically search for high proper motion objects using the newly released catalogue of epochal WISE data in order to identify new brown dwarf candidates in the solar neighborhood, estimate their spectral types, distances and spatial velocities.}
   {We used recently released unTimely catalogue of epochal detections in unWISE coadds to search for objects with high proper motions using simple motion detection algorithm, \new{combined with machine learning based artefact rejection routine.} This method was used to identify objects with proper motions exceeding approximately 0.3 arcseconds per year. The identified objects were then cross-referenced with data from other large-scale sky surveys to further analyze their characteristics.
   }
   {The search yielded \new{21,885} moving objects with significant proper motions, \new{258} of which had not been previously published. \new{All except 6 of them are compatible with being ultracool dwarfs}. Among these, at least \new{33} were identified as \new{most promising} new T dwarf candidates, with estimated distances closer than about 40 parsecs, \new{and effective temperatures less than 1300 K}.
   }
   {}


   \maketitle
%

\section{Introduction}
\label{sec:introduction}

Brown dwarfs are substellar objects with masses insufficient to start and maintain stable hydrogen fusion, which causes them to cool over time. They were theoretically predicted \citep{Kumar1963,Hayashi1963} and then later discovered \citep{Rebolo1995, 1995Natur.378..463N}. Since then, the search  for new brown dwarfs, and systematic study of known ones is a major problem in stellar astrophysics.

\new{According to the studies \citep{Muzic2017}, brown dwarfs are estimated to constitute up to 25\% of the stellar and substellar objects in the Galaxy. Recent studies \citep{kirk2024} offer more conservative estimates, suggesting a star-to-brown dwarf ratio of 4:1.
Searching for brown dwarfs is crucial for understanding the full spectrum of stellar and substellar populations in the Galaxy. Brown dwarfs are needed for advancing our understanding of the initial mass function (IMF), a key astrophysical tool for exploring star formation, the evolution of galaxies, and the distribution of stellar masses \citep{2003PASP..115..763C, 2013pss5.book..115K, 2024ApJ...974..222R, 2024ApJS..271...55K, 2024arXiv241007311K}. Finding new brown dwarfs enhances our understanding of Galactic kinematics and expands our understanding of stellar populations \citep{2010ApJS..190..100K, smith_2014, 2024AJ....168..159L}. Additionally, investigating the atmospheres of brown dwarfs is vital for unraveling the complex processes that influence their temperature, composition, and cloud formation \citep{Khandrika2013ASF, Burningham2017}, ultimately providing a deeper understanding of similar mechanisms in the atmospheres of gas giant exoplanets \citep{Marley2015, Charnay2018, Tan2019, Tremblin2019}.}

The search for brown dwarfs has evolved significantly over the years, building upon techniques initially developed for finding late-type M dwarfs in photographic surveys. Among these, photometric selection has emerged as the dominant strategy for searching for brown dwarfs in wide-field surveys. The method's effectiveness relies on the key difference in spectral characteristics between the target objects and the majority of the background population. This distinction allows one to identify the specific region of color space that characterises primary brown dwarf population.

The searches  based on color selection method and characterisation of brown dwarfs were performed using DENIS \citep{1997A&A...327L..25D, 2008MNRAS.383..831P}, 2MASS \citep{1999ApJ...519..802K}, SDSS \citep{2000AJ....119..928F, 2006AJ....131.2722C}, UKIDSS \citep{2013MNRAS.430.1171D, 2013MNRAS.433..457B}, CFDBS \citep{2008A&A...482..961D}, WISE \citep{2011ApJS..197...19K} and most recently DES \citep{2019MNRAS.489.5301C, des_dwarfs} surveys. In many cases, cross-matching different catalogues are useful. For example, the 2MASS passbands alone are not sufficient to distinguish from stars of other classes, although they help in classifying between subclasses of brown dwarfs. Furthermore, by combining optical data from Pan-STARRS with WISE data, researchers have successfully targeted LT transition objects that are typically hard to detect in near-infrared surveys \citep{2015ApJ...814..118B, 2018ApJS..234....1B}.

\new{Such selection, however, has rarely been perfect, even with a combination of various surveys. Photometric samples are often contaminated with extraneous objects, necessitating spectroscopic confirmation to establish reliable datasets. While the search conducted by \citet{2010AJ....139.1808S} utilized the SDSS DR7 spectroscopic database and resulted in the discovery of 210 new brown dwarfs, revealing that existing color selection methods are biased towards redder brown dwarfs, it is important to acknowledge that this was not the only spectroscopic search for brown dwarfs. Subsequent studies, such as \citet{2022A&A...660A..38W}, who explored UCD spectra in LAMOST data, and \citet{2024ApJS..274...40H}, who performed a comprehensive analysis of UCD candidates in APOGEE DR17, have also contributed significantly to the field. Additionally, \citet{2024AJ....168..179L} utilized JWST/NIRISS to identify brown dwarfs in NGC 1333.}

\new{Multiple brown dwarfs have been recently discovered at kiloparsec distances, see the results from JADES/CEERS surveys~\cite{2024ApJ...964...66H}, and UNCOVER~\cite{2024ApJ...962..177B} survey. Furthermore, brown dwarfs have been recently discovered in closest globular clusters~\cite{2024ApJ...971...65G}.
However, most brown dwarfs are discovered in the nearest solar neighbourhood, so their proper motions should be large, compared to the majority of the field stars. Combined with typically suboptimal angular resolution of infrared instruments, it significantly complicates the association of measurements between different datasets acquired at different epochs, leading to predominantly slower moving objects being detected using color selection methods. }


To overcome this bias, uniform data sets with dense multi-epoch coverage are needed.
The multi-epoch data from 2MASS survey \citep{2010ApJS..190..100K} and from AllWISE Motion Survey \citep{2014ApJ...783..122K} with a spectroscopic follow-up have revealed previously unknown LT dwarfs and subdwarfs. The large-scale search for rapidly moving brown dwarfs is nowadays enabled by the CatWISE2020 \citep{catwise2020} catalogue that reports both infrared color and proper motions for nearly two billions of objects over the whole sky. However, as the object detection in CatWISE2020 is performed on the co-added images combining all unWISE epochs, it is not optimal for finding rapidly moving stars, and prone to numerous spurious detections due to aggressive deblending applied to the data. Thus, the projects like Backyard Worlds  have to utilize citizen science approach for visual inspection of many candidate sources in order to find actual moving objects \citep{byw_first, byw_comoving, byw_wide}, \new{as well as advanced machine learning based selection of brown dwarf candidates \citep{byw_coolneighbors, byw_ml_catwise}.}

On the other hand, recent release of unTimely \citep{untimely} catalogue of epochal  detections in individual unWISE coadds opens the possibility of direct search for most rapidly moving objects. Here we describe such a search that we performed on $W2$ band data in order to specifically find new candidates for nearby \new{ultracool dwarfs, i.e. late M stars and brown dwarfs}. In Section~\ref{sec:methods} we describe the algorithm and data we used, as well as the moving objects we detected. Section~\ref{sec:discussion} contains the \new{comparison of our method performance to other brown dwarf search efforts,  reviews the properties and possible classification of new ultracool dwarf candidates we found}. Finally, Section~\ref{sec:conclusions} concludes the work.


\section{Data and analysis}
\label{sec:methods}

The Wide-field Infrared Survey Explorer, or WISE \citep{wise}, is a space telescope that performed all-sky survey initially in four infrared bands ($W1$ at 3.4$\mu$m, $W2$ at 4.6$\mu$m, $W3$ at 12$\mu$m and $W4$ at 22$\mu$m), and since the coolant depletion in late 2010 -- in two shorter-wavelengths bands only. It repeatedly scanned the sky in great circles near a solar elongation of 90$\degr$, typically observing a given region of the sky over a period of 1 d every 6 months, with denser coverage closer to the ecliptic poles. unWISE coadds combine the exposures from the same sky regions into a series of six-monthly visits, with an average of typical 12 exposures per band per visit \citep{unwise,unwise_2018}. These coadds are both sharper and deeper than original data used for producing initial AllWISE catalogue \citep{allwise}, and are the basis of unWISE \citep{unwise_cat} and CatWISE2020 \citep{catwise2020} catalogues greatly improving over it. However, these catalogues still did not provide the detections or measurements for the objects at individual epochs.

In order to systematically search for cool rapidly moving objects in WISE data we used recently released unTimely catalogue \citep{untimely} which is the result of uniform analysis of individual epochal unWISE coadds \citep{unwise,unwise_2018} using \texttt{crowdsource} crowded-field photometry package \citep{crowdsource,crowdsource_ascl}. It is currently distributed as a set of FITS tables each corresponding to the objects detected in one of two filters ($W1$ or $W2$) in individual epochal coadds for one of 18240 sky tiles. Typically every tile is covered with 15 to 17 unWISE coadds spanning approximately 11 years between Jan 2010 and Dec 2020 in every filter, but some of them (284 tiles, located close to ecliptic poles) have significantly more, up to two hundred, epochs spanning the same time interval \citep{unwise_2018}.

Epochal catalogues are independent, and not cross-matched together, both between different epoch and between different filters in a single epoch. Therefore, as we are primarily interested in detecting cool objects, we processed only the catalogues in $W2$ (4.6$\mu$m) band. Moreover, to ensure the consistency of depths of catalogues from individual epochs, we excluded from the analysis the epochs with total number of contributing exposures (\texttt{N\_EXP} field in the unTimely metadata table) less than a half of the median value for that tile.
We applied basic quality cuts to the objects from epochal catalogues
by selecting only the detections from primary parts of unWISE coadds (\texttt{primary}==1)
having flux measured with signal to noise ratio of at least \new{5}. Additionally, we require the detections to
have sufficiently large
fraction of flux inside the PSF that comes from this object (\texttt{fracflux}>0.5) to ensure that the detections are not subject to significant crowding, and thus have reliable astrometry and (less relevant for us) photometry.
After initial experimenting we decided not to perform any filtering on ``quality factor'' (\texttt{qf}) field or flags derived either from unWISE coadds themselves (\texttt{flags\_unwise}) or from \texttt{crowdsource} processing (\texttt{flags\_info}) as they prevent most brighter objects from being detected, while not reducing significantly the amount of typical artefacts among the candidates.
\new{The positional accuracy of positions in epochal catalogues strongly depends on the source brightness. Our initial experiments demonstrated that \texttt{crowdsource}-measured uncertainties (\texttt{dx} and \texttt{dy} fields) adequately reflect it for objects fainter than 8th mag, while brighter ones show significantly larger scatter, most probably due to being saturated. Therefore, we assumed constant positional uncertainty of 1$''$ for the latters. Moreover, we detected systematic bias between positions measured during forward and backward scans. To compensate for it, we added an extra positional error of 0.15$''$ to all measurement uncertainties.}

\subsection{Motion detection algorithm}
\label{sec:algorithm}

\begin{figure}
\centering
\includegraphics[width=\columnwidth]{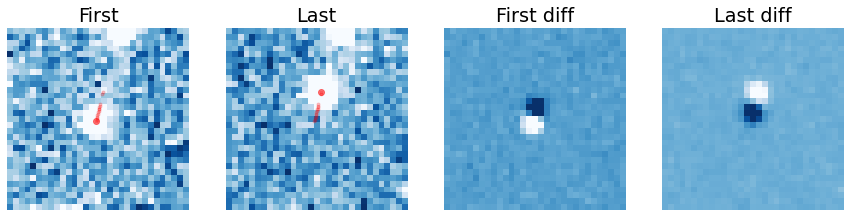}
\includegraphics[width=\columnwidth]{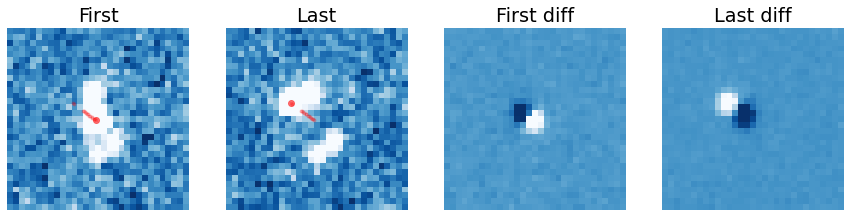}
\includegraphics[width=\columnwidth]{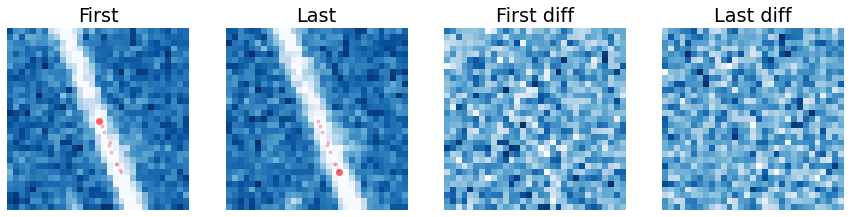}
\includegraphics[width=\columnwidth]{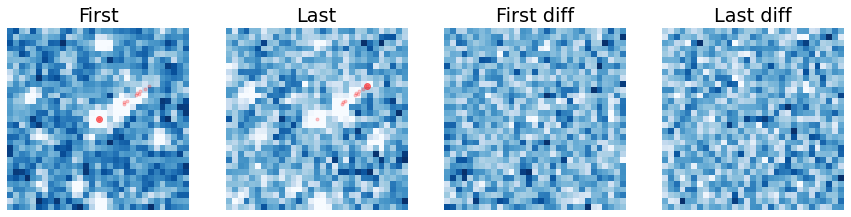}
\caption{Examples of images used to visually check the moving object candidates selected by the algorithm as described in Section~\ref{sec:algorithm}. Each row displays the cutouts from the first and last epochs in unWISE coadds, and the corresponding differences between the images and median of all available epochs. All cutouts are centered on the position of the object from the first epoch (so that it is in the center in the first and third columns). Red dots mark the positions of the objects from all available epochs, with the larger circle corresponding to the current epoch of the image.\\
First and second rows display the actual moving objects, for an isolated and crowded fields, while the third one shows an artefact due to bright star spike, and the last -- spurious candidate due to linking together individual detections of several stationary objects located roughly along the line.
}
\label{fig:diff}
\end{figure}

\begin{table}
\caption{\new{Features used for the machine learning classifier as described in Section~\ref{sec:algorithm}, and their relative importance for the classification at the final iteration.}}
\label{tab:features}
\centering
\footnotesize

\begin{tabular}{cp{5cm}c}
\hline\hline
Name & Description & Importance \\
\hline

\texttt{pm} & Estimated proper motion & 0.01 \\
\texttt{e\_pm} & Statistical error of estimated proper motion & 0.17 \\
\texttt{N} & Number of epochs with detections & 0.10 \\
\texttt{corr\_30} & Pearson correlation coefficient of the coordinates vs time for all points within 3$''$ from expected position at different epochs & 0.23 \\
\texttt{corr} & Pearson correlation coefficient of the coordinates vs time for all points within 3$\sigma$ brightness dependent coordinate uncertainty from expected position at different epochs & 0.16 \\
\texttt{magstd} & Actual RMS of individual magnitude measurements in $W2$ band & 0.14 \\
\texttt{magerr} & Mean error of individual magnitude measurements in $W2$ band & 0.06 \\
\texttt{mag} & Mean magnitude in W2 band & 0.05 \\
\texttt{magchi2} & Reduced $\chi^{2}$ of individual magnitude measurements in $W2$ band & 0.02 \\
\texttt{b} & Galactic latitude & 0.01 \\
\texttt{NNNN1} & Number of detections with 1 or more catalogue points not belonging to the track within its 3$\sigma$ positional uncertainty & 0.02 \\
\texttt{NNNN3} & the same for 3 or more points & 0.01 \\
\texttt{NNNN2} & the same for 2 or more points & 0.01 \\
\texttt{NNNN4} & the same for 4 or more points & 0.001 \\
\texttt{NNNN5} & the same for 5 or more points & 0.001 \\

\hline
\end{tabular}

\end{table}

\begin{figure}
\centering
\includegraphics[width=\columnwidth]{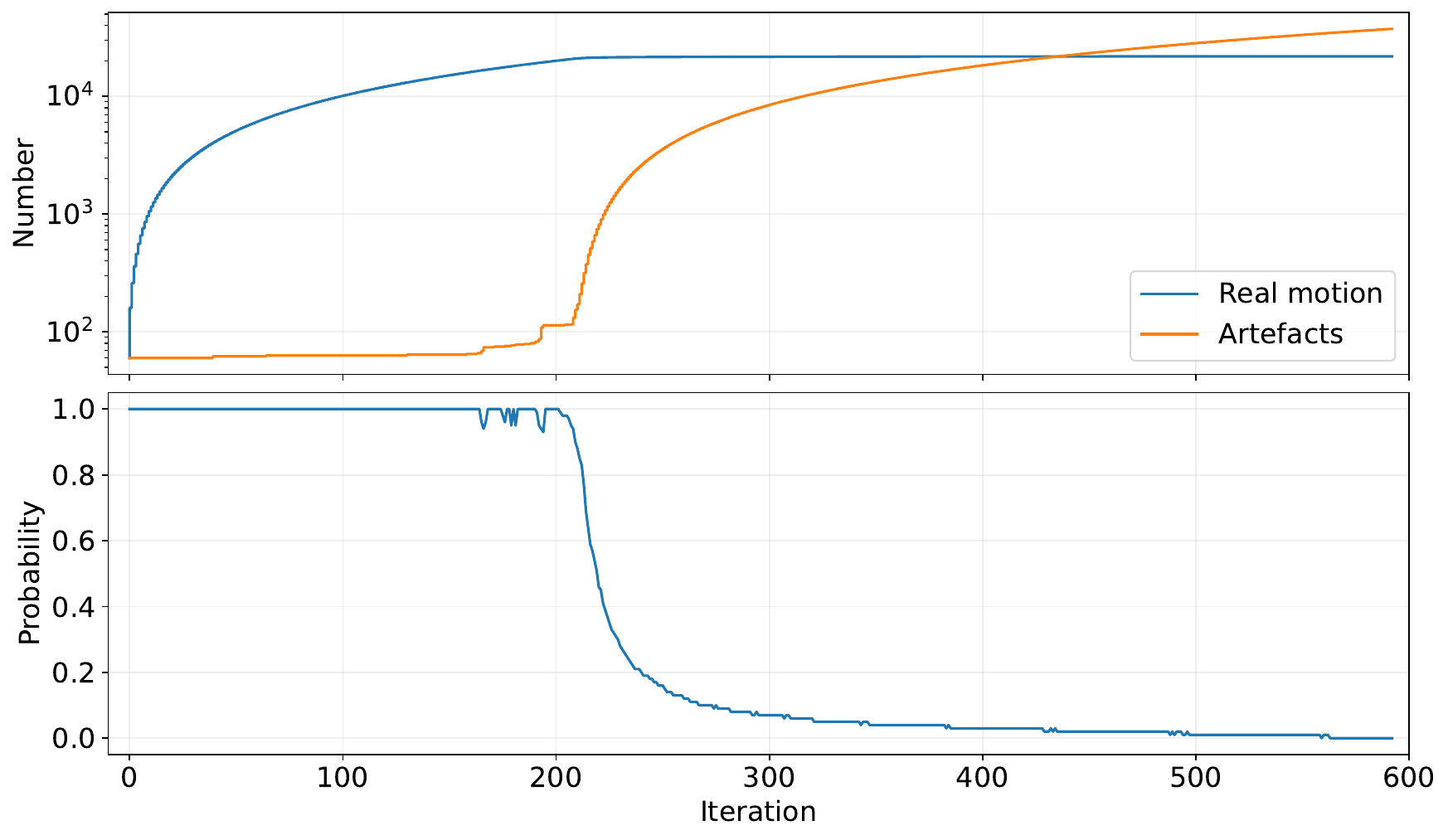}
\caption{\new{The progress of machine learning based classification of initial candidates, as described in Section~\ref{sec:algorithm}. Upper panel -- number of visually vetted real moving objects and artefacts at each iteration, from the pool of 1,271,921 initial candidates. Lower panel -- threshold probability corresponding to 100 previously unclassified candidates with highest score produced by Random Forest binary classifier trained at each iteration. These candidates are then visually checked and used for improving the classifier on next iteration. The procedure stops when threshold probability falls to zero (the classifier stops producing new candidates for visual inspection).}}
\label{fig:iterations}
\end{figure}

\begin{figure}
\centering
\includegraphics[width=\columnwidth]{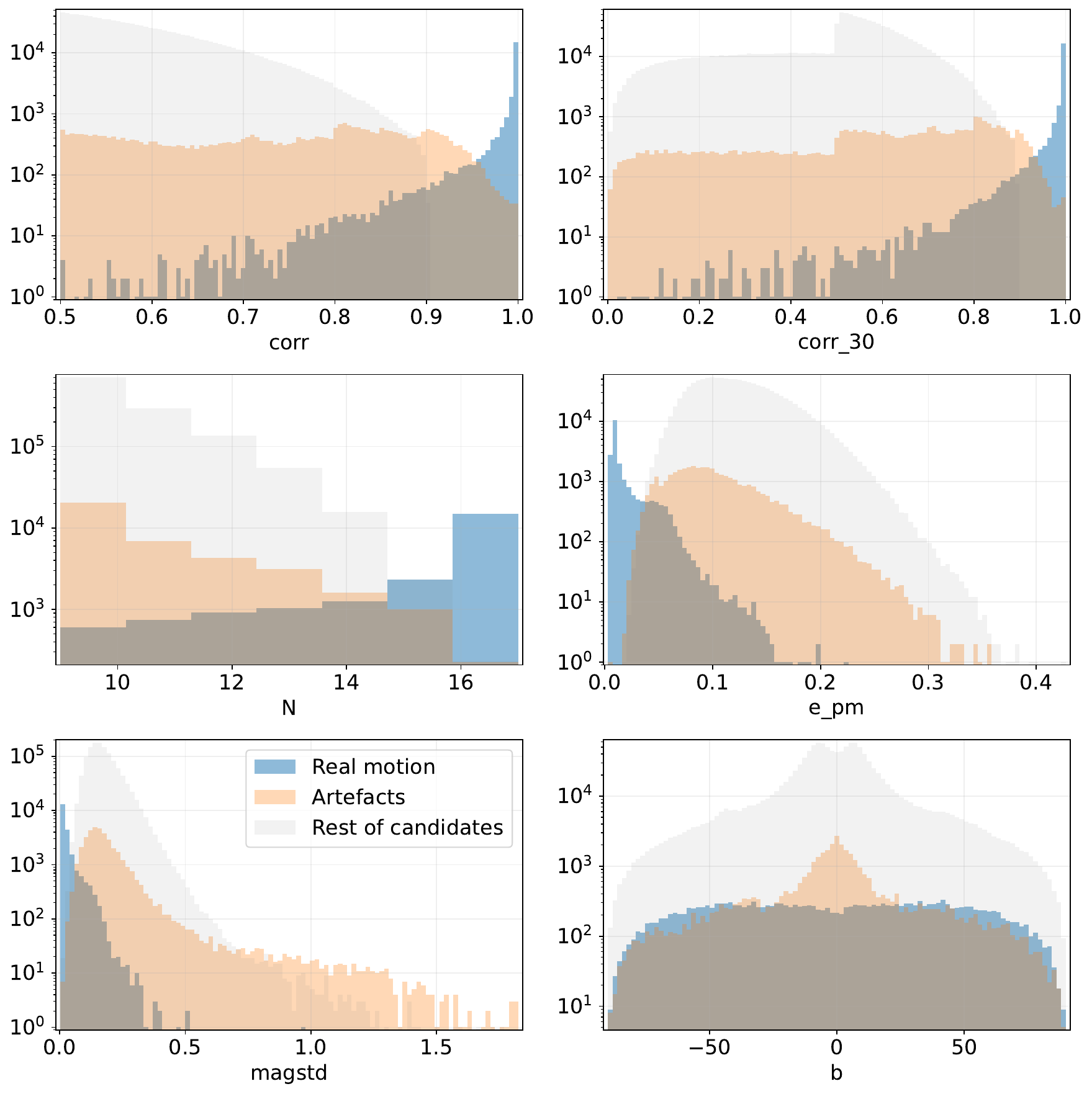}
\caption{\new{Histograms of several features, as defined in Table~\ref{tab:features}, for visually vetted real moving objects, visually vetted artefacts, and the rest of candidates rejected by the iterative machine learning classification as described in Section~\ref{sec:algorithm}. The rest of the candidates never got a high score from the classifier, and their distribution is similar to the one of visually vetted artefacts.}
}
\label{fig:hists}
\end{figure}

\new{Actual motion detection is performed using the algorithm similar to the one implemented in e.g. \citet{canfind} and based on finding clusters of points located along the lines in a three-dimensional space of sky coordinates and time. However, in contrast to optical data in \citet{canfind}, it is enormously complicated by significant crowding and relatively low positional accuracy of the catalogue points, comparable to the actual proper motion of most objects over ten years time span.}

After applying various quality cuts as described above we analyzed the sets of all objects from epochal catalogues for every tile in order to find candidate moving objects. \new{To do so, we converted their sky coordinates to standard coordinates relative to the tile center, and proceeded to detection of linear structures in resulting three-dimensional space (two angular coordinates plus time) using a} simple  ``opportunistic'' algorithm that consider all pairs of points from different epochs that satisfy the criteria of being separated by at least \new{3}$''$ (slightly more than 1 pixel in unWISE coadds, and slightly less than half of full-width at half maximum (FWHM) of its point-spread function (PSF)) but no more than \new{50}$''$ (for reducing the computational complexity). This criterion ensures significant motion of the candidates, effectively placing a lower limit on the detectable proper motion of about \new{0.3}$''$/year. \new{In order to lower the contamination due to stationary objects, we excluded the pairs where either of the points has other detections within 1$''$ from it in more than $60\%$ of epochs with the same brightness, within measurement uncertainties.}
Then, for every pair, we formed a candidate track, computed its expected positions for all the epochs available for that tile, and then selected the objects from corresponding epochs (this time without applying any additional criteria on top of initial quality cuts) that are closer than \new{3$\sigma$ positional uncertainty} to these positions. We rejected the candidate tracks that do not have \new{measurements} in more than 60\%  of the epochs, corresponding to at least 10 epochs in typical 16-epochal tile \new{without low-quality epochs. We also required the modulo of Pearson Correlation Coefficient of at least one angular coordinate with time to be at least 0.5.
For the tracks passing these criteria, we computed the proper motion by performing robust linear regression of their standard coordinates versus time using RANSAC\citep{ransac} algorithm as implemented in \textsc{scikit-learn}\citep{scikit-learn} Python package in order to mitigate the influence of spurious associations with stationary objects, and then rejected the tracks with relative accuracy of proper motion estimation less than 0.33, thus ensuring that the candidate motion is significant. At this stage we also computed various statistical properties for the points belonging to the tracks after RANSAC filtering, as well as nearby catalogue points from different epochs, to serve later as features for artefact rejection.}

The candidates proposed by the algorithm are prone to both genuine image-level artefacts, especially due to long spikes around bright stars, and spurious tracks formed from individual detections of stationary objects located approximately along a line, \new{especially for fainter candidates with larger positional uncertainties of individual detections}. We could not find a way to reliably filter out either kind of artefacts using the information available in the catalogue files, \new{and had to rely on the machine learning based filtering method that will be described in more details below, and finally on the visual vetting of the candidates} with a simple \new{Jupyter-based} \citep{jupyter} dashboard based on the WISEView \citep{unwise_2018,wiseview} cutout API developed in the frame of Backyard Worlds project \citep{byw_first}. The dashboard displays, for every candidate, the cutouts from the first and last epochs at the object position, and the corresponding differences between these images and the median image of all epochs, and allows easy selection of real moving objects by the characteristic dipoles in the latter images, as shown in Figure~\ref{fig:diff}. \new{The dashboard displays up to several hundreds of candidates in a single page along with interactive controls for their flagging as either artefacts or actual moving objects, with the ability to also directly check }
the epochal animations using the original WISEView web tool\footnote{Available at \url{http://byw.tools/wiseview}.} \new{in the unclear cases.}

\new{
The algorithm returned 1,271,921 initial candidates from all tiles of unTimely catalogue. Despite the visual vetting dashboard allowing a person to inspect thousands of the candidates in relatively small amount of time, this amount is still too much. Therefore we developed a simple iterative routine, akin to the one used by \citet{byw_ml_catwise}, for extracting ``true'' moving objects from them based on Random Forest\footnote{\new{We used Random Forest classifier implementation from \textsc{scikit-learn} \citep{scikit-learn}, with 100 estimators and automatically balanced class weights.}} \citep{randomforest} binary classifier and a set of features for the candidates extracted by the algorithm, as listed in Table~\ref{tab:features}. We bootstrapped the classifier by selecting and visually checking the candidates with 60 largest (which all appeared to be true moving objects) and 60 smallest (which all were artefacts) values of the correlation of coordinates with time (\texttt{corr} parameter), and training the Random Forest on these objects. We then applied the classifier to all candidates, selected a hundred ones with largest predicted probability of being true, and visually inspected them, then using derived labels to improve the classifier and repeat the whole process. It's progress is shown in Figure~\ref{fig:iterations}. We finished the process when the classifier stopped predicting new candidates with non-zero probabilities. At that point, we got 21,885 candidates visually vetted to be real moving objects, and 37,437 -- as various artefacts.
We characterized the performance of final classifier using \textit{Leave One Out} cross-validation -- by repeatedly excluding single points to serve as a test set, and training the model on the rest. After doing full cycle, we estimated mean precision of the classifier to be 99.6\%, and mean recall -- 96.5\%.
}

\new{Figure~\ref{fig:hists} shows the distribution of several features  for both real moving objects and artefacts, both visually vetted and rejected by the classifier. The amount of the latters significantly grows towards lower number of epochs, larger uncertainties of estimated proper motions, larger photometric scatter, as well as towards the Galactic plane with largest density of stationary objects.}

\subsection{Performance verification}
\label{sec:performance}

\begin{figure}
\centering
\includegraphics[width=\linewidth]{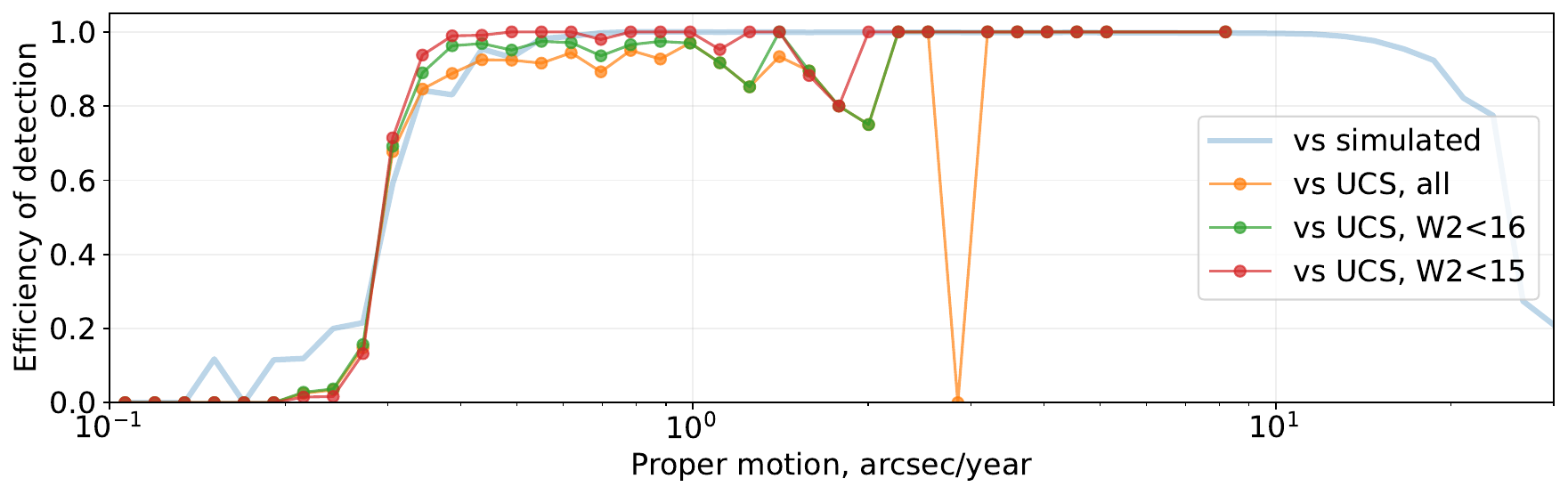}
\includegraphics[width=\linewidth]{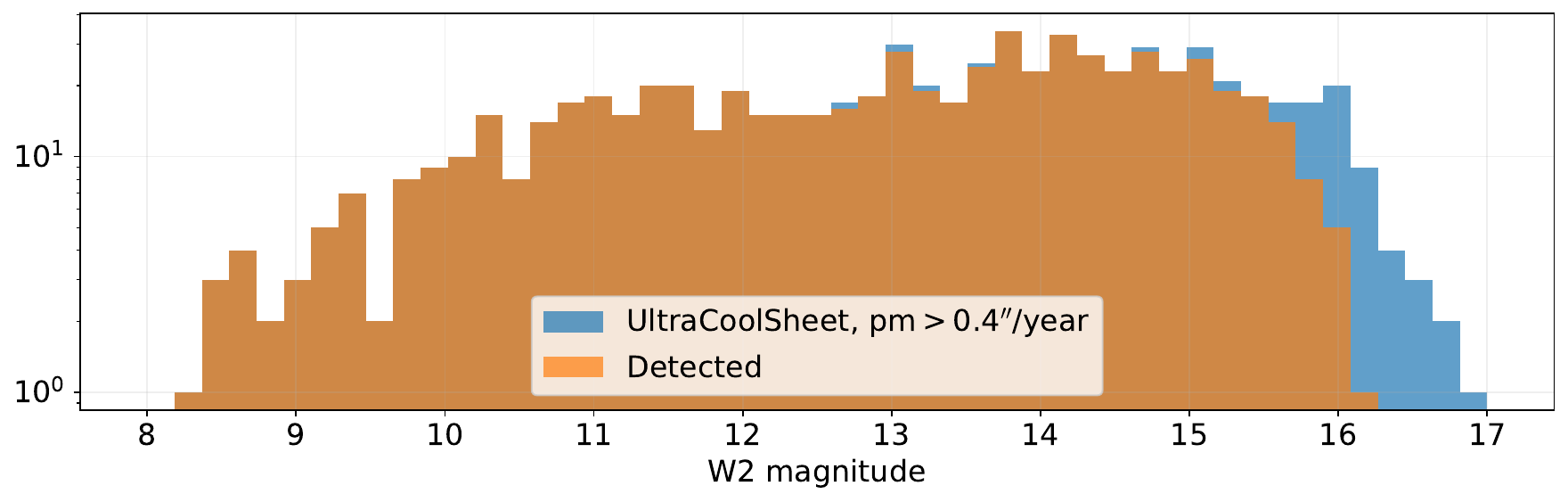}
\caption{\new{Upper panel -- efficiency of the moving object detection algorithm, estimated as described in Section~\ref{sec:performance}
by either injecting simulated entries into unTimely epoch catalogues (only detection \textit{per se} is characterized), or by comparing of the results with the list of known ultracool dwarfs from UltracoolSheet \citep{ucs} (both detection and machine learning based filtering are characterized) for different brightness in WISE $W2$ band.
As expected, the majority of objects faster than $\sim$0.3$''$/year are successfully being detected, with nearly 100\% efficiency for the objects brighter than $W2<15$, and expected slight performance loss for fainter ones due to epochal unTimely detection limit being worse than CatWISE2020 \citep{catwise2020} one.
The algorithm is equally efficient for the objects moving as fast as 10$''$/year, and still has expected efficiency of 82\% at 20$''$/year.
Lower panel -- histogram of the $W2$ magnitude for known UltracoolSheet objects moving faster than 0.4$''$/year, and its subset detected here.
}}
\label{fig:efficiency}
\end{figure}

\begin{figure}
\centering
\includegraphics[width=\linewidth]{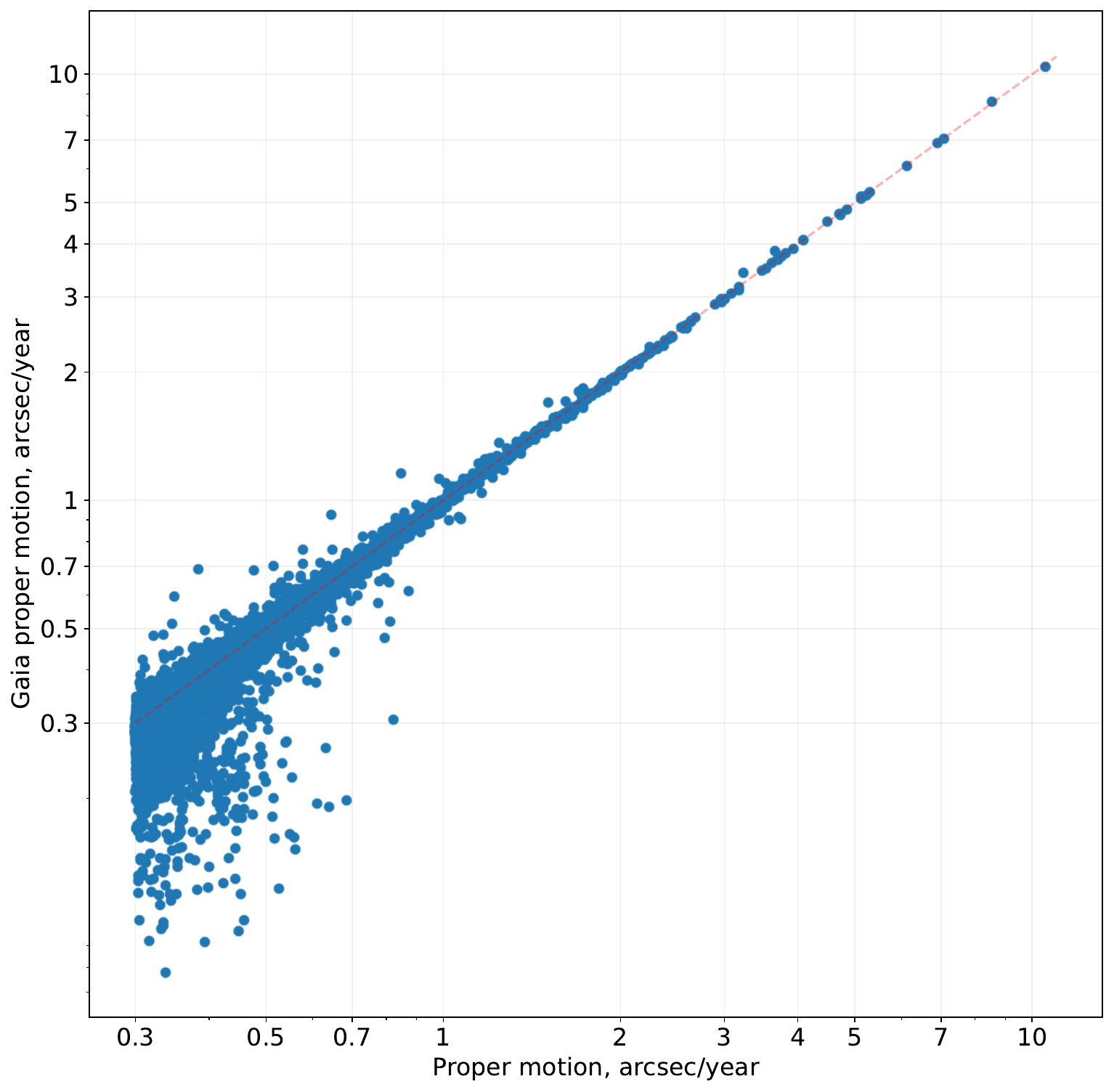}
\caption{Comparison of the proper motions measured in this work with values from Gaia DR3 \citep{gaiadr3}. The agreement with Gaia proper motions is sufficiently good, with the spread increasing towards smaller values as the track length in unTimely data becomes too small to permit accurate measurement.
}
\label{fig:pm_gaia}
\end{figure}

\new{As a result of iterative machine learning based filtering of initial candidates described above, we visually confirmed 21,885 moving objects among 59,322 candidates we manually inspected in total, giving 37\% overall efficiency of the algorithm as a whole.
In order to control the sensitivity of the algorithm to different values of proper motion we performed a run of the candidate extraction routine with additional injection into every tile of 100 simulated objects with random positions, velocities and brightness, and positional scatter corresponding to actual uncertainties for this brightness, as described above. We processed these simulated objects together with actual data, so that they have the same density of background distractor points as real moving objects, and checked the fraction of these events successfully recovered by the algorithm. As expected (see Figure~\ref{fig:efficiency}), the algorithm practically does not detect the objects moving slower than 0.3$''$/year, but recovers all of the ones moving as fast as 10$''$/year (corresponding to fastest known star), and also most of the ones up to 20$''$/year, thus proving that our choice of the algorithm parameters is sufficient for detecting even fastest of moving objects at stellar distances.
}

\new{As we simulated the data on catalogue level, we expect that this test does not include all selection effects affecting real data. Moreover, it does not include the machine learning classification step (as simulation does not realistically reproduce  all the data properties needed for computing the features). Therefore, we also validated the algorithm by cross-matching moving objects we detected with the UltracoolSheet \citep{ucs, ucs1, ucs2} compilative database of ultracool dwarfs\footnote{UltracoolSheet is available online at \url{http://bit.ly/UltracoolSheet}.} that contains the
most comprehensive list of known nearby cool objects.
We filtered it to only contain entries having WISE associations to ensure they are potentially visible, and checked the fraction that is actually detected by the algorithm and passed machine learning filtering, as described above. As shown in Figure~\ref{fig:efficiency}, the efficiency estimated this way is consistent with simulations for brighter objects ($W2<15$) and is slightly less for fainter ones due to unTimely epochal catalogues being not as deep as CatWISE2020 based on full unWISE coadds \citep{catwise2020}.}

\new{To assess the accuracy of the proper motion determination by the algorithm} we performed a cross-match of detected objects
with Gaia~DR3 catalogue \citep{gaiadr3} by propagating object positions to its J2016.0 epoch, using 2$''$ matching radius, and only considering Gaia objects with significant proper motions to avoid confusions with faint stationary background stars. This way we found the associations for \new{20555 (94\%)} of all candidates. \new{For 104 more candidates, mostly brighter ones, where Gaia~DR3 did not provide an association we found matching counterparts in Gaia~DR2 catalogue.
Figure~\ref{fig:pm_gaia} shows that the agreement of proper motions is sufficiently good and unbiased for its whole range, except for the slowest objects with shortest arcs where our measurements tend to be overestimated so that they pass the 0.3$''$/year algorithm threshold.}

We also cross-matched the candidates with CatWISE2020 \citep{catwise2020} catalogue by propagating their positions to J2015.4 (MJD=57170) epoch, using catalogue columns corresponding to proper motion corrected positions for that epoch, and conservatively assuming 3$''$ (slightly more than unWISE coadds pixel size) matching radius. While we do find the associations for \new{most candidates (except for 1052 which are most probably either too bright or located close to imaging artefacts)}, for significant part of them the association is not unique and contains, in addition to seemingly unbiased ones with comparable proper motions and fluxes, also the counterparts that are $\sim$1 magnitude fainter and have $\sim$1$''$/year greater velocity. We attribute these spurious associations to systematic deblending problems in the catalogue introducing artificial detections. \new{It also manifests in a huge number of bogus high proper motion objects in the catalogue, making its use problematic for searching for actual moving objects.
However, we tend to believe that photometric measurements there are superior to unTimely catalogue ones, and we will use these magnitudes for our objects in the analysis below.
}


\subsection{New high proper motion objects}
\label{sec:new}

\begin{figure}
\centering
\includegraphics[width=\linewidth]{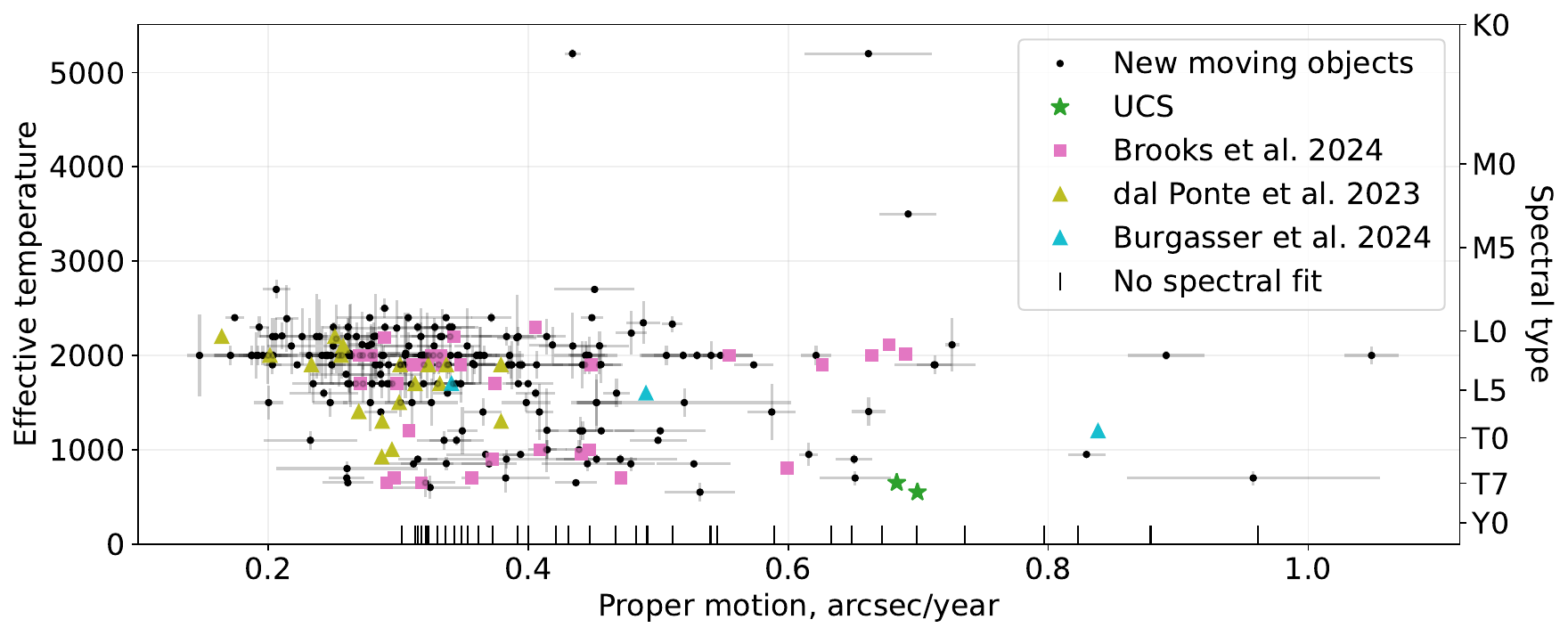}
\caption{\new{High proper motion objects detected in this work that do not have SIMBAD or Gaia DR3/DR2 associations. Proper motions are estimated using both unTimely data, and additional associations from several multiwavelength photometric catalogues. Effective temperatures are from the fit of these photometric data with the grid of BT-Settl  \citet{btsettl} model spectra, as described in Section~\ref{sec:new}.
Colored symbols represent known objects we also detected and characterized that are already published in UltracoolSheet \citep{ucs,ucs1,ucs2}, \citet{brooks_2024}, \citet{des_dwarfs}, or \citet{burg2024}, and thus excluded from the final list.
Black vertical ticks at the bottom mark the proper motions of new objects that do not have enough multiwavelength associations for spectral fitting.
Secondary vertical axis on the right shows the spectral types corresponding to the effective temperatures according to \citet{mamajek}.
}}
\label{fig:pm_teff}
\end{figure}

\begin{figure*}
\centering
\includegraphics[width=\columnwidth]{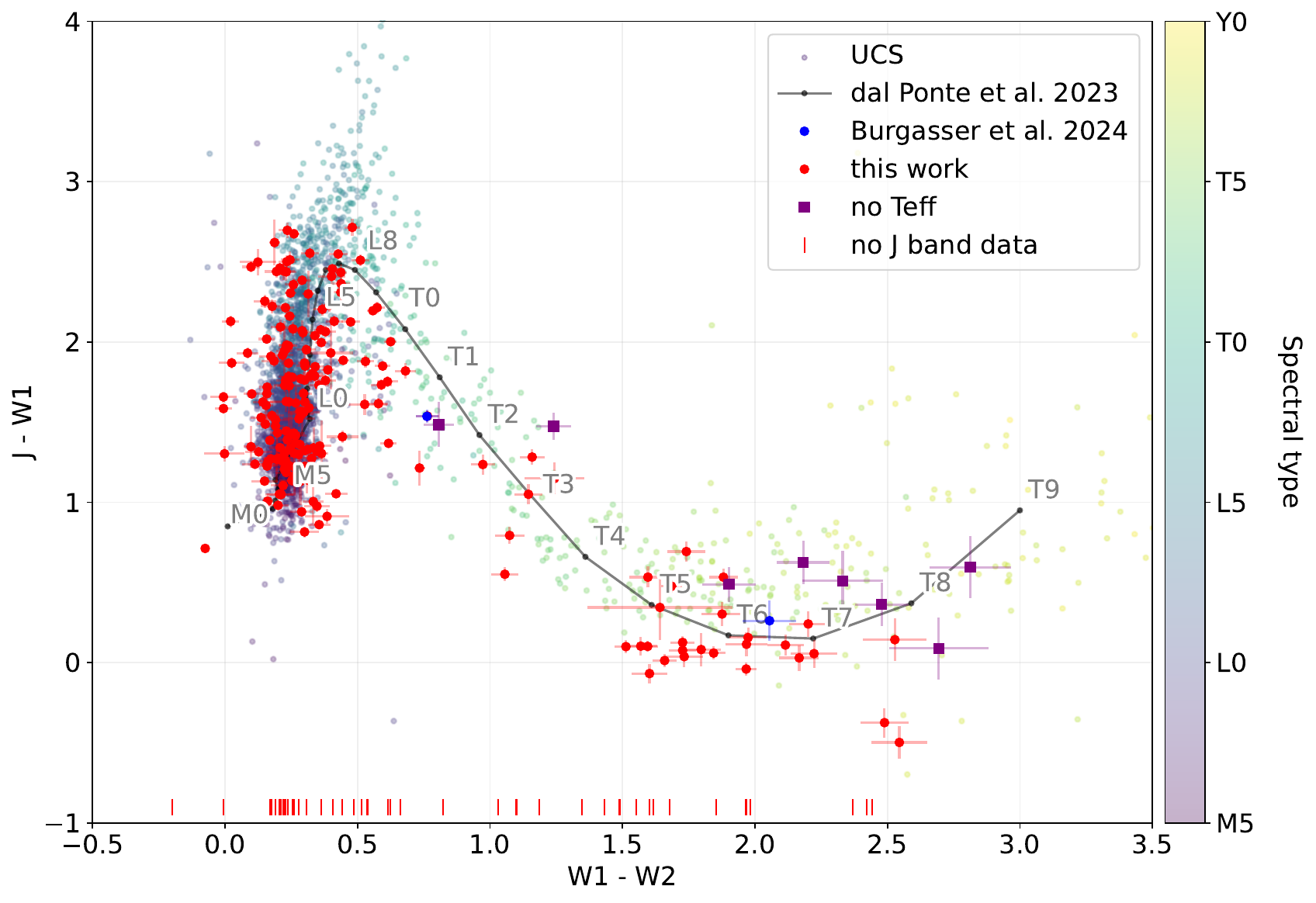}
\includegraphics[width=\columnwidth]{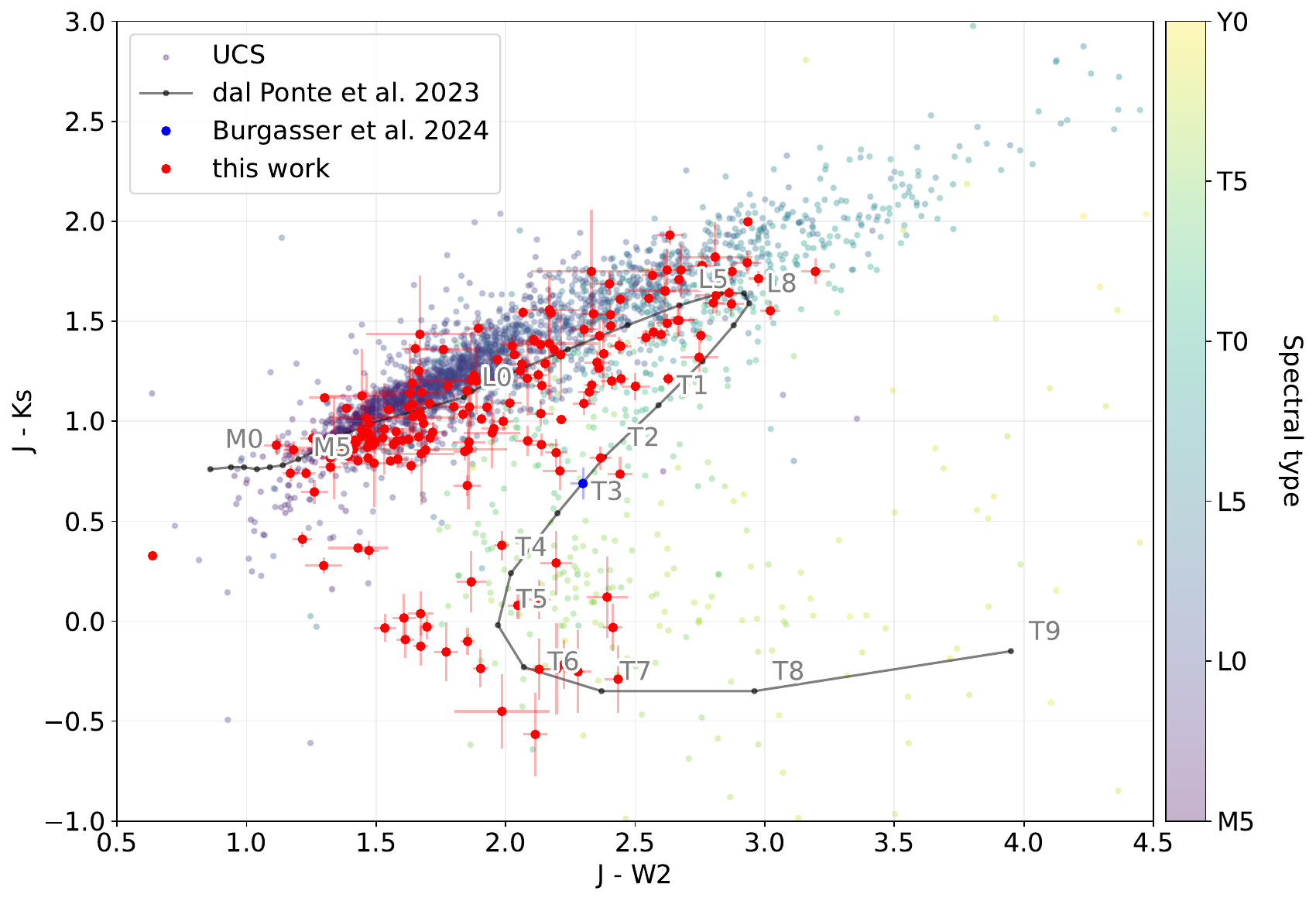}
\includegraphics[width=\columnwidth]{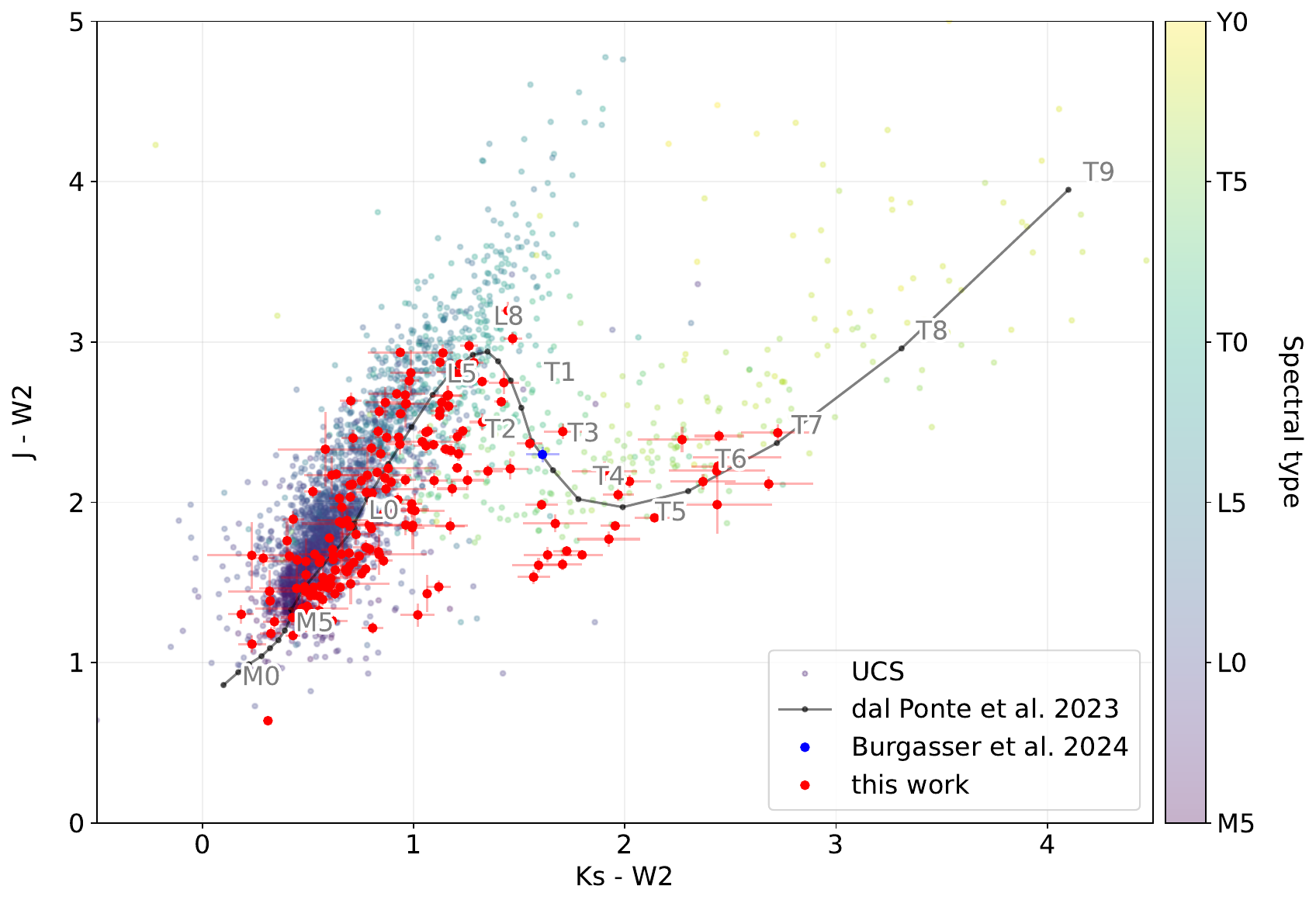}
\includegraphics[width=\columnwidth]{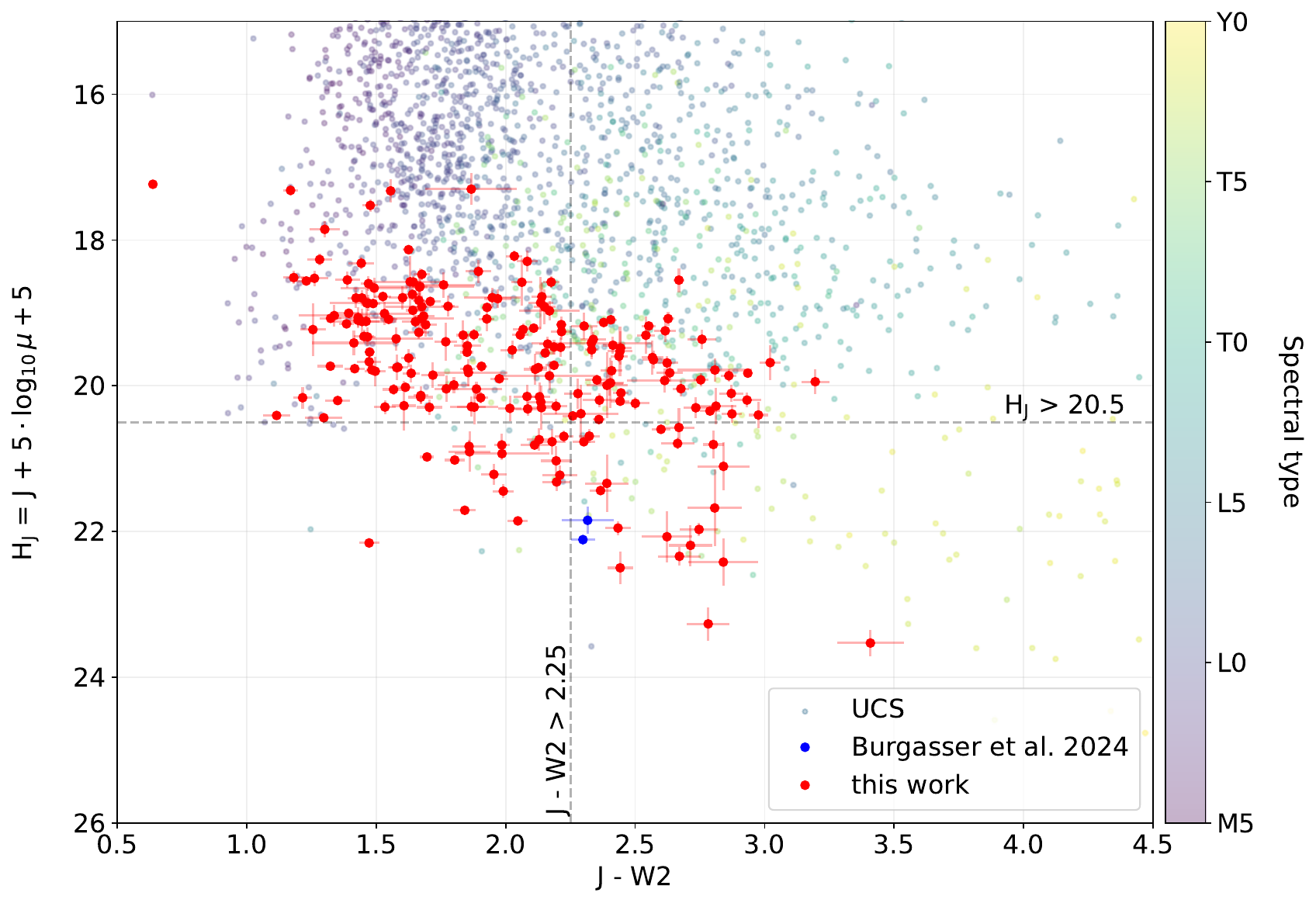}
\caption{Positions of the new moving objects we detected
(red circles with error bars) in \new{infrared two-color and reduced J magnitude} diagrams for the known ultracool dwarfs from UltracoolSheet  \citep{ucs}. The colors of known objects correspond to their spectral class according to the color bars to the right of the plots.
\new{Blue points show spectroscopically confirmed T subdwarfs from \citet{burg2024} that we also detected.}
Black line marks the template colors of ultracool dwarfs according to \citet{des_dwarfs}.
\new{Red vertical ticks at the bottom of the left panel mark the WISE color of objects that do not have associated $J$ band measurements. Purple points also mark the objects where we did not get enough multiwavelength data for VOSA spectral fit, but $J$ band data are still available.
Black dashed lines in the bottom right panel mark the criteria for selection of candidate T subdwarfs in \citet{burg2024}.
}
}
\label{fig:colors}
\end{figure*}

\begin{figure}
\centering
\includegraphics[width=\linewidth]{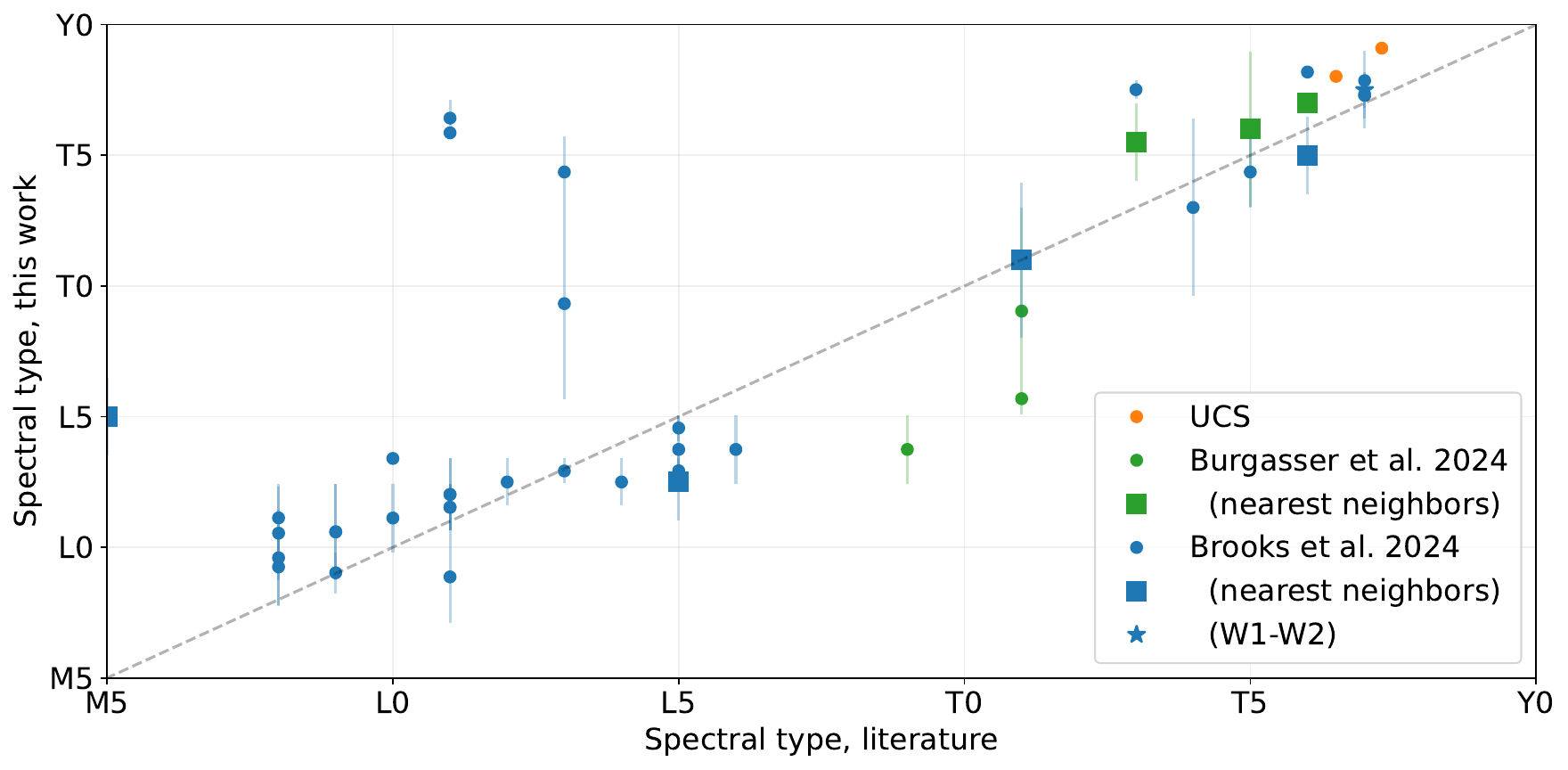}
\caption{\new{Comparison between published spectral types for the objects that do not have neither SIMBAD nor Gaia associations but are present in either UltracoolSheet \citep{ucs}, \citet{brooks_2024}, or \citet{burg2024}, with our best estimates done using VOSA spectral fit based on BT-Settl model grid (circles), using nearest UltracoolSheet neighbors in the two-color plots akin to Figure~\ref{fig:colors} (squares), or by using UltracoolSheet objects with similar $W1-W2$ colors (stars). Error bars correspond to 96\% (2$\sigma$) confidence limits of the corresponding estimations.
Two significant outliers in the upper left corner are CWISE~J220134.02-041609.5 and CWISE~J230252.77-131458.0 that both have very large $\chi^2$ values in \citet{brooks_2024} indicating bad fits there, so that we consider our late type estimations for them more correct. VOSA effective temperatures for three objects from \citet{burg2024} agree well with published values, while their spectral types are underestimated (green circles below the equality line) due to differences in the temperature to spectral type mapping for subdwarfs from \citet{mamajek} ones we used.
}}
\label{fig:spt_comp}
\end{figure}

We checked for associated SIMBAD \citep{simbad} database entries for \new{all 21,885 objects with high proper motions that we detected}. As SIMBAD is highly heterogeneous and does not generally provide the epochs for the coordinates listed there, we, for every candidate, computed a number of trial positions propagated to the epochs from J1990.0 till J2021.0 with a step small enough to accommodate for candidate' proper motion, and looked for SIMBAD entries within 5$''$ from these positions. \new{We only considered relevant types of objects, thus excluding e.g. extragalactic sources, or objects from spectral ranges other than optical and infrared. This way, we found the associations for all objects except for 515 ones.
Among them, 192 have counterparts with measured proper motions in Gaia DR3 or DR2 catalogues. After initial analysis of their colors we did not find among them candidates consistent with spectral types later than L0, so, like \citet{brooks_2024}, we decided to exclude them from the analysis but use them later for assessing the reliability of our estimations (see Section~\ref{sec:discussion}. Among the rest, 2 objects are listed in UltracoolSheet \citep{ucs,ucs1,ucs2} compilative database of ultracool dwarfs, 19 are in the list of ultracool dwarf candidates from DES 6-year data \citep{des_dwarfs}, 38 are recently found in unWISE images by \citep{brooks_2024}, and 6 more were recently reported in a spectroscopic study of cool subdwarfs from Backyard Worlds by \citet{burg2024}.
We excluded them all from our final list (again, keeping them for cross-checking the performance of our analysis), leaving us with 258 new moving objects.
}

\new{Three of them also have no counterparts in CatWISE2020 \citep{catwise2020} catalogue.
For them, we used mean values\footnote{\new{Our analysis shows that in crowded regions unTimely epochal magnitudes are highly correlated with the \texttt{fracflux} parameter describing the fraction of the flux inside PSF footprint coming from the object itself. While regressing for it and then extrapolating to \texttt{fracflux}=1, i.e. fully uncrowded case, looks like proper routine to get unbiased object magnitude from unTimely data, direct comparison with CatWISE2020 shows that such estimates are significantly less stable and more biased than just mean values.}}
of unTimely epochal measurements for $W1$ and $W2$ band magnitudes, while for the rest we directly used CatWISE2020 magnitudes as more accurate, especially for $W1$ band, and less prone to be biased due to variable crowding.
}

\new{We collected additional photometric data for these objects by cross-matching their expected positions at corresponding epochs with several large-scale optical and IR catalogues that have measurement epochs for their detections -- 2MASS point source catalogue \citep{2mass}, the catalogue of Pan-STARRS DR2 epochal (``warp'') detections \citep{ps1}, VISTA Hemisphere Survey \citep{vhs}, UKIRT Hemisphere Survey \citep{uhs} epochal detections, SkyMapper DR4 \citep{skymapper_dr4} epochal detections, Sloan Digital Sky Survey DR16 \citep{sdss_dr16}, DENIS \citep{denis},  NOIRLab Source Catalogue DR2 \citep{nsc_dr2}, and DELVE DR2 \citep{delve}, with the search radius large enough to accommodate for both expected position and catalogue uncertainties. We then visually inspected all the associations to ensure that, together with epochal unTimely detections, they form both consistent trajectory and consistent spectral energy distribution. To assess the latter, we fit the measurements with the grid of spectra for BT-Settl model atmospheres \citet{btsettl} using VO SED Analyzer (VOSA\footnote{VO SED Analyzer is available online at \url{http://svo2.cab.inta-csic.es/svo/theory/vosa/}}) \citep{vosa}. After excluding spurious associations, we got both improved proper motion estimates, sets of photometric measurements in optical and IR bands, and the effective temperatures from the spectral fits for 223 objects, as shown in Figure~\ref{fig:pm_teff}. The spectral type may then be estimated by interpolating the values from Mamajek table of mean colors and effective temperatures for dwarf stars\footnote{\new{``A Modern Mean Dwarf Stellar Color and Effective Temperature Sequence'' version 2022.04.16 is available online at Eric Mamajek website at \url{https://www.pas.rochester.edu/~emamajek/EEM_dwarf_UBVIJHK_colors_Teff.txt}} \citep{mamajek}.}
}

\new{Two-color diagrams constructed from measurements in different IR bands are shown in Figure~\ref{fig:colors}, together with the positions and spectral classes of known ultracool dwarfs from UltracoolSheet, and template ultracool dwarf colors from \citet{des_dwarfs}.
}

\new{For 38 of new moving objects we did not get enough (at least 5) multiwavelength points for VOSA model fit, so no temperature estimates are available for them. However, for some of them the data from shorter wavelengths are still available -- see e.g. left panel of Figure~\ref{fig:colors} --  and may be used to assess their spectral type using a number of nearest neighbors from UltracoolSheet with known spectral types in corresponding two-color spaces (in the order of $J-W1$ vs $W1-W2$, $y-W1$ vs $W1-W2$, $z-W1$ vs $W1-W2$ and $i-W1$ vs $W1-W2$). The scatter of spectral types among nearest neighbors will then define the error for the estimation.
This way we estimated spectral types for 27 of these objects.
For the rest, we may use single $W1-W2$ color as a regressor for spectral type, again using UltracoolSheet data, and averaging among all its entries with the color laying within the corresponding error bars for our objects. Again, the scatter of catalogue spectral types in the subset will define the error of estimated spectral type.
We do not expect this method to be accurate for $W1-W2<0.5$, but for redder objects (L9 and later types) we consider such estimation reasonable.
Figure~\ref{fig:spt_comp} shows the comparison between  published spectral types for the objects that our dataset without SIMBAD or Gaia associations has in common with UltracoolSheet \citep{ucs}, \citet{brooks_2024}, or \citet{burg2024}, with our best estimates made as described above. It shows a good agreement, especially for later types, except for several outliers that we believe to be mis-classifications in \citet{brooks_2024} due to spurious measurements from shorter wavelengths included in the template fits there.
}

\new{Table~\ref{tab:summary} summarizes the parameters of 258 new moving objects that we found, both astrometric (positions and proper motions) and photometric (from multi-wavelength associations), along with inferred effective temperatures and spectral types. 214 of the objects have estimated temperatures colder than 2700K, and thus are ultracool dwarfs. For 38 more, spectral types estimated from infrared colors are later than M7, thus also making them ultracool dwarf candidates.
}


\begin{table*}
\caption{\new{New high proper motion objects detected in this work, listed in the order of decreasing proper motion. Only the objects with reasonably late spectral types (Teff$ < $1300K or $W1-W2 > 0.8$, roughly corresponding to T0 spectral type), and a subset of columns, are shown, while the complete table is available online.
The table is published at \url{https://doi.org/10.5281/zenodo.14651236}.
}}
\label{tab:summary}
\centering
\small
\begin{tabular}{cccccccccccc}
\hline\hline
CWISE & RA & Dec & pm & W2 & W1-W2 & J & Ks & Teff & SpT & Dist & V$_{\mbox{tan}}$ \\
 & J2016.0 & J2016.0 & $''$/year &  &  &  &  & K & & pc & km/s \\

\hline
J011816.64-723917.9 & 19.5683 & -72.6550 & 0.96 & 15.39 & 1.62 &  &  &   & T4--T6\tablefootmark{2} & 31 $\pm$ 5 & 141 $\pm$ 33 \\
J105018.24-683054.9 & 162.5751 & -68.5153 & 0.96 & 15.15 & 2.20 & 17.59 &  & 700--850 & T7--T8  & 23 $\pm$ 4 & 104 $\pm$ 22 \\
J214516.34+235253.0 & 326.3177 & 23.8812 & 0.88 & 15.77 & 2.69 & 18.55 &  &   & T6--T8\tablefootmark{1} & 29 $\pm$ 7 & 120 $\pm$ 33 \\
J214706.01-540407.6 & 326.7749 & -54.0688 & 0.83 & 15.21 & 2.55 & 17.26 & 17.18 & 950--1000 & T6  & 29 $\pm$ 3 & 115 $\pm$ 14 \\
J140048.92+253759.8 & 210.2037 & 25.6333 & 0.82 & 15.55 & 2.81 & 18.95 &  &   & T7--T9\tablefootmark{1} & 24 $\pm$ 6 & 95 $\pm$ 25 \\
J233228.09-301831.8 & 353.1171 & -30.3089 & 0.74 & 15.30 & 1.68 &  &  &   & T3--T6\tablefootmark{1} & 32 $\pm$ 5 & 112 $\pm$ 16 \\
J045456.23-742522.6 & 73.7343 & -74.4229 & 0.70 & 15.48 & 2.42 &  &  &   & T4--Y0\tablefootmark{2} & 25 $\pm$ 8 & 83 $\pm$ 26 \\
J114557.19+673327.5 & 176.4881 & 67.5576 & 0.67 & 15.88 & 1.03 &  &  &   & T2\tablefootmark{1} & 56 $\pm$ 6 & 179 $\pm$ 24 \\
J020149.83-050629.6 & 30.4577 & -5.1084 & 0.65 & 15.45 & 1.74 & 17.88 & 18.17 & 700--850 & T7--T8  & 26 $\pm$ 5 & 82 $\pm$ 15 \\
J024641.43-634329.2 & 41.6727 & -63.7248 & 0.65 & 15.21 & 1.60 & 16.91 & 16.94 & 900--1000 & T6  & 27 $\pm$ 4 & 85 $\pm$ 12 \\
J184526.47+754829.7 & 281.3602 & 75.8084 & 0.63 & 15.12 & 1.55 &  &  &   & T4--T6\tablefootmark{1} & 27 $\pm$ 5 & 82 $\pm$ 15 \\
J095855.00+540305.4 & 149.7291 & 54.0515 & 0.62 & 15.27 & 1.73 & 17.07 &  & 950--1200 & T3--T6  & 32 $\pm$ 4 & 92 $\pm$ 13 \\
J221213.11-332147.9 & 333.0546 & -33.3634 & 0.59 & 15.28 & 1.35 &  &  &   & T3--T5\tablefootmark{1} & 32 $\pm$ 6 & 90 $\pm$ 20 \\
J212303.09+301042.3 & 320.7628 & 30.1786 & 0.54 & 15.82 & 1.24 & 18.53 &  &   & T2--T4\tablefootmark{1} & 47 $\pm$ 8 & 121 $\pm$ 26 \\
J045646.17-530953.7 & 74.1924 & -53.1649 & 0.53 & 16.04 & 2.53 & 18.71 &  & 550--750 & T8--T9  & 29 $\pm$ 6 & 72 $\pm$ 16 \\
J145540.52+424812.0 & 223.9186 & 42.8034 & 0.53 & 15.51 & 2.17 & 17.71 & 17.95 & 850 & T7  & 30 $\pm$ 4 & 75 $\pm$ 12 \\
J125933.55+560059.2 & 194.8900 & 56.0166 & 0.51 & 16.04 & 2.33 & 18.88 &  &   & T6--T8\tablefootmark{1} & 33 $\pm$ 8 & 79 $\pm$ 23 \\
J030426.27-612844.7 & 46.1102 & -61.4791 & 0.50 & 15.33 & 0.62 & 17.31 & 16.93 & 1200 & T3  & 42 $\pm$ 8 & 100 $\pm$ 21 \\
J170259.75-090457.4 & 255.7489 & -9.0826 & 0.50 & 15.26 & 1.60 & 16.80 & 16.83 & 1100 & T5  & 30 $\pm$ 4 & 71 $\pm$ 9 \\
J175502.15+502858.0 & 268.7590 & 50.4827 & 0.49 & 16.09 & 2.44 &  &  &   & T4--Y0\tablefootmark{2} & 33 $\pm$ 10 & 77 $\pm$ 25 \\
J230906.24-152344.7 & 347.2761 & -15.3958 & 0.48 & 15.55 & 1.64 & 17.53 & 17.98 & 850--1000 & T6--T7  & 32 $\pm$ 5 & 73 $\pm$ 12 \\
J150336.92-694649.9 & 225.9036 & -69.7808 & 0.47 & 15.47 & 1.14 & 17.67 & 17.37 & 900--1000 & T6  & 31 $\pm$ 4 & 69 $\pm$ 11 \\
J071705.85-580823.3 & 109.2745 & -58.1396 & 0.47 & 15.24 & 1.96 &  &  &   & T5--T7\tablefootmark{2} & 25 $\pm$ 6 & 56 $\pm$ 13 \\
J071250.77-804800.9 & 108.2113 & -80.8002 & 0.46 & 15.17 & 1.66 & 16.84 & 16.96 & 1200 & T3  & 39 $\pm$ 8 & 85 $\pm$ 17 \\
J141403.72-171531.1 & 213.5155 & -17.2587 & 0.45 & 15.33 & 1.60 & 17.46 & 17.70 & 900--1000 & T6  & 29 $\pm$ 4 & 62 $\pm$ 10 \\
J081447.10+375829.7 & 123.6966 & 37.9751 & 0.45 & 14.91 & 1.10 &  &  &   & T1--T6\tablefootmark{2} & 25 $\pm$ 5 & 54 $\pm$ 13 \\
J133053.00-135919.1 & 202.7207 & -13.9887 & 0.45 & 15.34 & 1.88 & 17.52 &  & 851--1000 & T6--T7  & 29 $\pm$ 4 & 62 $\pm$ 10 \\
J004247.39-092349.4 & 10.6976 & -9.3970 & 0.44 & 15.24 & 0.76 & 17.54 & 16.45 & 1200--1395 & L9--T3  & 44 $\pm$ 7 & 93 $\pm$ 15 \\
J152313.35+533729.3 & 230.8056 & 53.6249 & 0.44 & 15.80 & 0.97 & 18.01 & 17.26 & 1200--1500 & L7--T3  & 60 $\pm$ 11 & 126 $\pm$ 24 \\
J141045.69-184500.8 & 212.6903 & -18.7504 & 0.44 & 14.71 & 1.73 & 16.56 & 16.66 & 1000--1200 & T3--T6  & 24 $\pm$ 3 & 51 $\pm$ 7 \\
J210733.23-622313.5 & 316.8886 & -62.3871 & 0.44 & 15.27 & 2.12 & 17.49 & 17.71 & 650 & T8  & 20 $\pm$ 3 & 41 $\pm$ 6 \\
J122941.10+474840.9 & 187.4213 & 47.8113 & 0.43 & 15.70 & 2.18 & 18.51 &  &   & T6--T8\tablefootmark{1} & 28 $\pm$ 7 & 57 $\pm$ 20 \\
J074140.90-694204.6 & 115.4209 & -69.7014 & 0.42 & 16.08 & 2.37 &  &  &   & T4--Y0\tablefootmark{2} & 33 $\pm$ 10 & 66 $\pm$ 21 \\
J031050.72-014103.3 & 47.7113 & -1.6844 & 0.41 & 15.40 & 1.57 & 17.07 & 17.03 & 1000--1200 & T3--T6  & 34 $\pm$ 5 & 66 $\pm$ 10 \\
J153922.90+441204.1 & 234.8455 & 44.2013 & 0.41 & 15.32 & 1.51 & 16.94 & 17.03 & 1000--1200 & T3--T6  & 32 $\pm$ 5 & 64 $\pm$ 9 \\
J141334.34+394207.4 & 213.3934 & 39.7020 & 0.40 & 15.94 & 1.90 & 18.33 &  &   & T5--T7\tablefootmark{1} & 35 $\pm$ 8 & 66 $\pm$ 19 \\
J045228.61-614041.3 & 73.1193 & -61.6781 & 0.39 & 15.28 & 1.84 & 17.19 & 17.42 & 950--1000 & T6  & 30 $\pm$ 4 & 56 $\pm$ 7 \\
J170103.70-363030.8 & 255.2655 & -36.5083 & 0.39 & 12.97 & 1.19 &  &  &   & T2--T5\tablefootmark{1} & 13 $\pm$ 2 & 23 $\pm$ 7 \\
J232520.76-764428.6 & 351.3373 & -76.7413 & 0.38 & 15.75 & 1.06 & 17.36 & 17.34 & 900--1100 & T5--T6  & 35 $\pm$ 5 & 64 $\pm$ 13 \\
J000604.71-190552.1 & 1.5197 & -19.0978 & 0.38 & 15.11 & 1.97 & 17.24 & 17.13 & 700--1000 & T6--T8  & 24 $\pm$ 5 & 43 $\pm$ 10 \\
J103304.78-035005.8 & 158.2699 & -3.8350 & 0.37 & 15.12 & 1.97 &  &  & 850--900 & T6--T7  & 25 $\pm$ 4 & 44 $\pm$ 7 \\
J094056.33-220743.7 & 145.2347 & -22.1287 & 0.37 & 15.45 & 1.73 & 17.22 & 17.37 & 950--1000 & T6  & 33 $\pm$ 4 & 57 $\pm$ 8 \\
J104104.55+782854.1 & 160.2688 & 78.4817 & 0.36 & 14.97 & 1.60 &  &  &   & T4--T6\tablefootmark{1} & 25 $\pm$ 4 & 44 $\pm$ 8 \\
J165024.50+622752.4 & 252.6022 & 62.4646 & 0.34 & 15.53 & 1.49 &  &  & 1100--1200 & T3--T5  & 36 $\pm$ 5 & 58 $\pm$ 9 \\
J194423.76+725402.3 & 296.0977 & 72.9007 & 0.34 & 15.45 & 1.85 &  &  &   & T5--T7\tablefootmark{2} & 28 $\pm$ 6 & 46 $\pm$ 11 \\
J233736.51-820951.7 & 354.4025 & -82.1644 & 0.34 & 15.78 & 1.80 & 17.66 &  & 853--1000 & T6--T7  & 36 $\pm$ 5 & 57 $\pm$ 10 \\
J171100.61+300617.1 & 257.7526 & 30.1047 & 0.34 & 15.79 & 1.07 & 17.66 & 17.46 & 1100--1300 & T0--T5  & 45 $\pm$ 11 & 71 $\pm$ 18 \\
J225625.60-273812.7 & 344.1067 & -27.6369 & 0.32 & 15.27 & 2.22 & 17.55 & 17.80 & 600--850 & T7--T8  & 22 $\pm$ 5 & 34 $\pm$ 8 \\
J215757.62-324339.7 & 329.4902 & -32.7277 & 0.32 & 14.62 & 0.82 &  &  &   & T1--T3\tablefootmark{1} & 30 $\pm$ 4 & 45 $\pm$ 9 \\
J161542.13+105557.8 & 243.9256 & 10.9328 & 0.32 & 15.53 & 1.97 & 17.62 &  & 650--950 & T6--T8  & 25 $\pm$ 6 & 38 $\pm$ 9 \\
J004424.68+195404.6 & 11.1030 & 19.9014 & 0.32 & 15.46 & 1.43 &  &  &   & T4--T6\tablefootmark{2} & 32 $\pm$ 6 & 48 $\pm$ 13 \\
J092245.41-835022.4 & 140.6898 & -83.8396 & 0.32 & 15.75 & 2.48 & 18.59 &  &   & T5--T8\tablefootmark{1} & 32 $\pm$ 7 & 48 $\pm$ 13 \\
J073205.97-582812.8 & 113.0249 & -58.4703 & 0.32 & 15.64 & 1.10 &  &  &   & T1--T6\tablefootmark{2} & 36 $\pm$ 8 & 53 $\pm$ 15 \\
J064952.51+593801.5 & 102.4688 & 59.6337 & 0.32 & 14.77 & 1.69 & 16.94 &  & 900 & T6  & 21 $\pm$ 3 & 32 $\pm$ 5 \\
J142336.58-391352.5 & 215.9025 & -39.2312 & 0.31 & 15.61 & 0.81 & 17.91 &  &   & T1--T3\tablefootmark{1} & 47 $\pm$ 6 & 69 $\pm$ 15 \\
J115016.18-310908.0 & 177.5673 & -31.1523 & 0.31 & 15.19 & 2.49 & 17.30 & 17.87 & 850 & T7  & 26 $\pm$ 4 & 38 $\pm$ 7 \\
J150950.62+701543.3 & 227.4609 & 70.2621 & 0.30 & 15.64 & 1.98 &  &  &   & T5--T7\tablefootmark{1} & 31 $\pm$ 7 & 44 $\pm$ 12 \\
J154802.28+565802.2 & 237.0094 & 56.9673 & 0.26 & 15.07 & 1.97 & 17.00 &  & 650 & T8  & 18 $\pm$ 3 & 22 $\pm$ 4 \\
J090705.67-675106.6 & 136.7744 & -67.8516 & 0.26 & 15.52 & 1.24 & 17.91 & 17.79 & 800--950 & T6--T7  & 27 $\pm$ 5 & 34 $\pm$ 9 \\
J094641.48-083945.4 & 146.6728 & -8.6626 & 0.26 & 14.95 & 1.88 & 17.37 & 17.40 & 700--800 & T7--T8  & 20 $\pm$ 3 & 24 $\pm$ 4 \\
J111748.85-181410.4 & 169.4535 & -18.2362 & 0.23 & 15.25 & 1.16 & 17.69 & 16.96 & 1100--1300 & T0--T5  & 35 $\pm$ 8 & 38 $\pm$ 11 \\
\hline
\end{tabular}

\tablefoot{\\
\tablefoottext{1} Spectral type estimated as a median spectral type of closest objects from UltraCoolSheet \citep{ucs} in two-color diagrams akin to Figure~\ref{fig:colors}. \\
\tablefoottext{2} Spectral type estimated from $W1-W2$ color and a table of colors from \citet{mamajek}. \\
}

\end{table*}


\section{Discussion}
\label{sec:discussion}

\begin{figure}
\centering
\includegraphics[width=\linewidth]{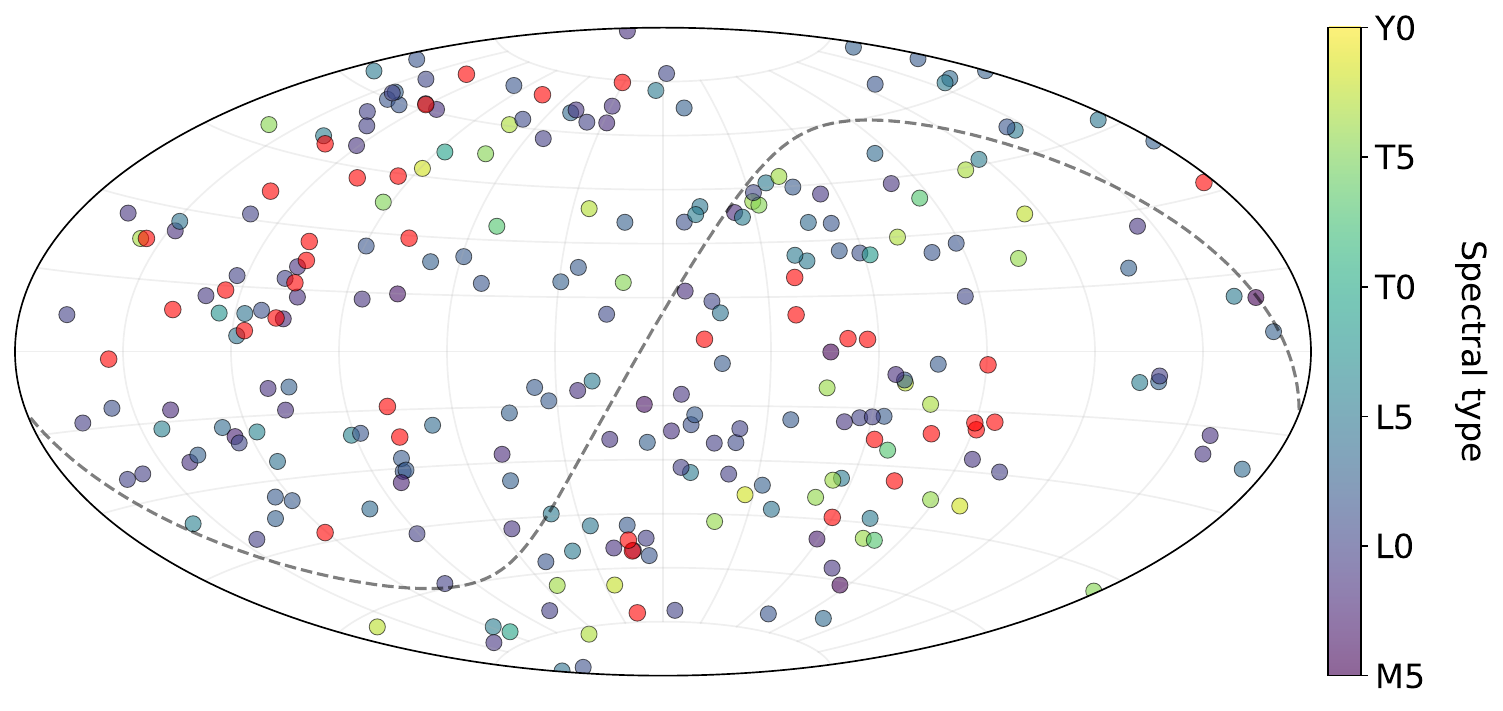}
\caption{\new{Positions of the detected moving objects in Galactic coordinates, with estimated spectral type represented by the marker color. The objects with no such estimates are colored in red.
Ecliptic plane is shown with black dashed line.}
}
\label{fig:galactic}
\end{figure}

\begin{figure}
\centering
\includegraphics[width=0.49\columnwidth]{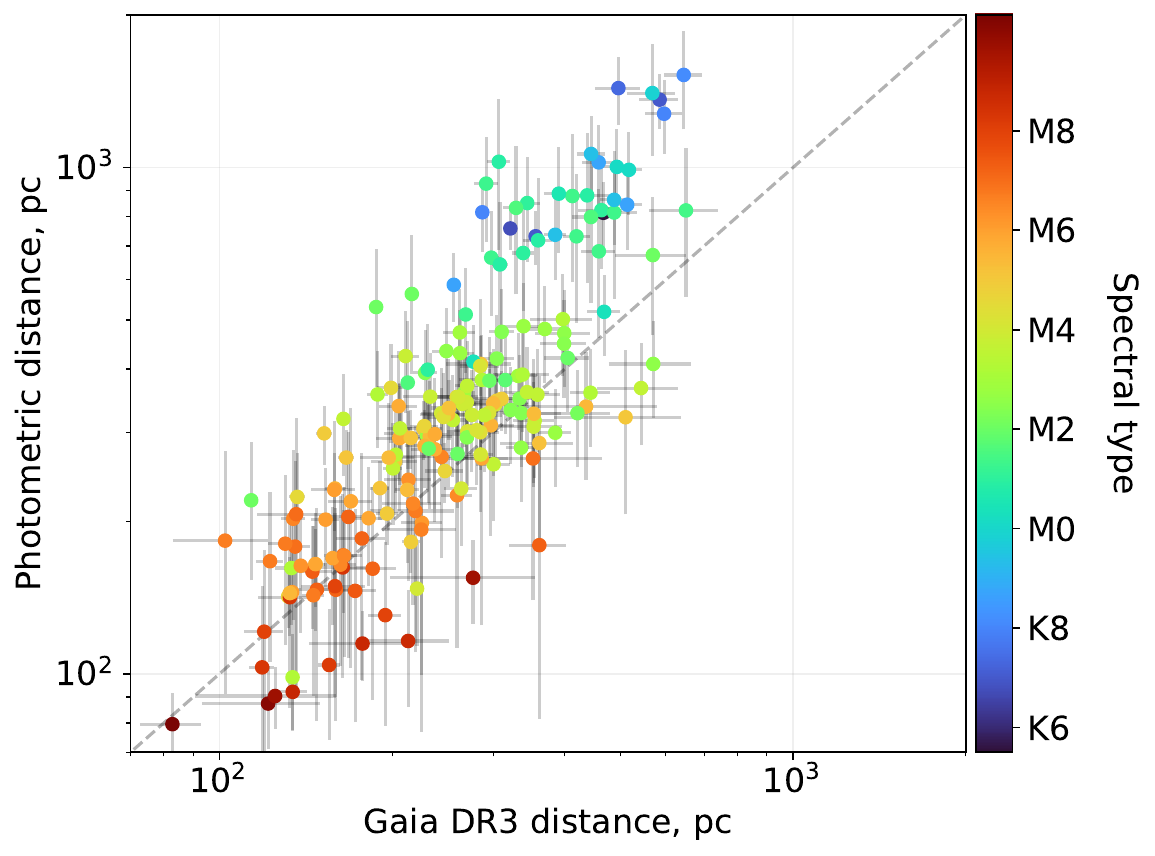}
\includegraphics[width=0.49\columnwidth]{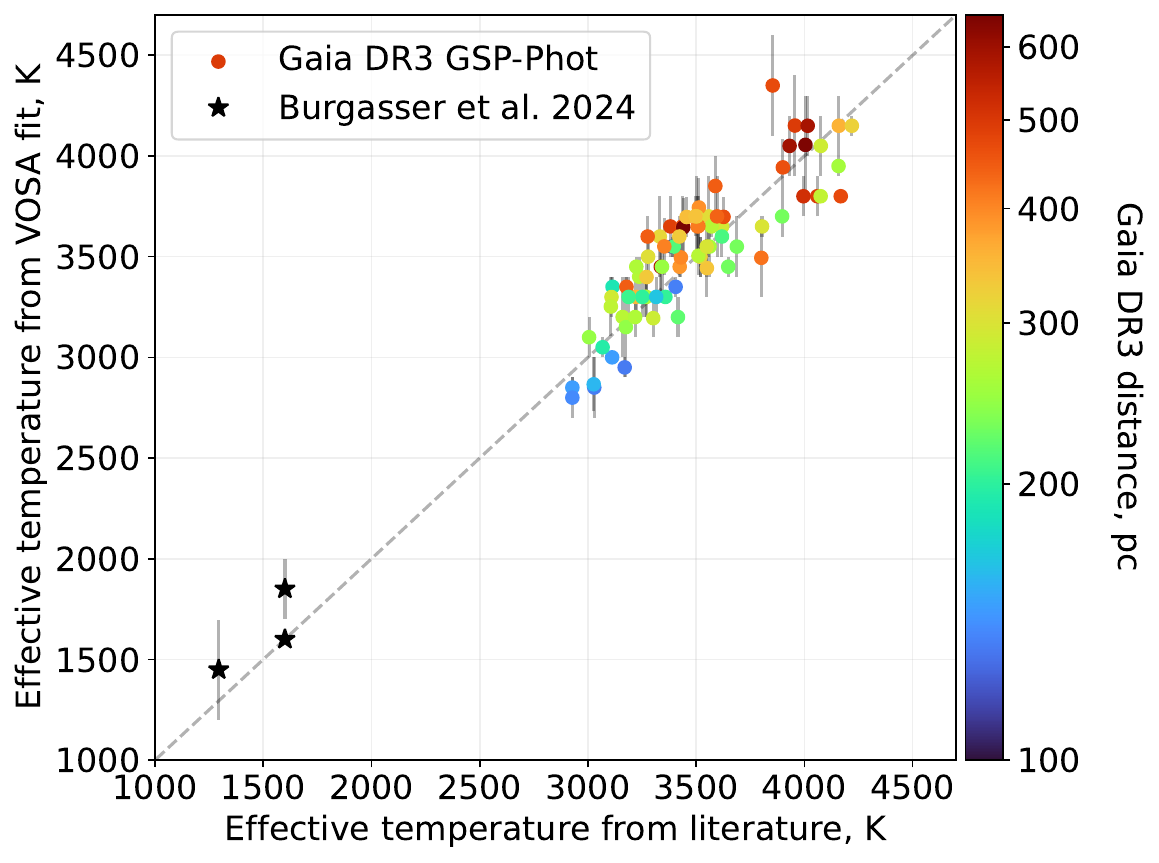}
\caption{\new{Left panel -- comparison of the Gaia DR3/DR2 distances for the moving objects lacking SIMBAD associations with photometric distances derived in Section~\ref{sec:discussion}. The distances estimated using interpolation of $M_W2$ absolute magnitudes are in good agreement with astrometric ones for the objects colder than approximately M3 spectral type, and are significantly overestimated for earlier types.
Right panel -- comparison of effective temperatures for the same objects estimated by Gaia DR3 GSP-Phot pipeline \citep{gaia_gspphot}, with the ones derived in this work using grid of BT-Settl model spectra implemented in VOSA\citep{vosa}. Colors of the dots show the distances from Gaia DR3. Overall agreement between VOSA and GSP-Phot estimations are sufficiently good.
Additionally, black stars represent the subdwarfs from \citet{burg2024} with effective temperatures derived from NIR spectra that overlap with our sample. Our estimated temperatures are also in good agreement with them.
}}
\label{fig:dist_gaia}
\end{figure}

\begin{figure*}
\centering
\includegraphics[width=\columnwidth]{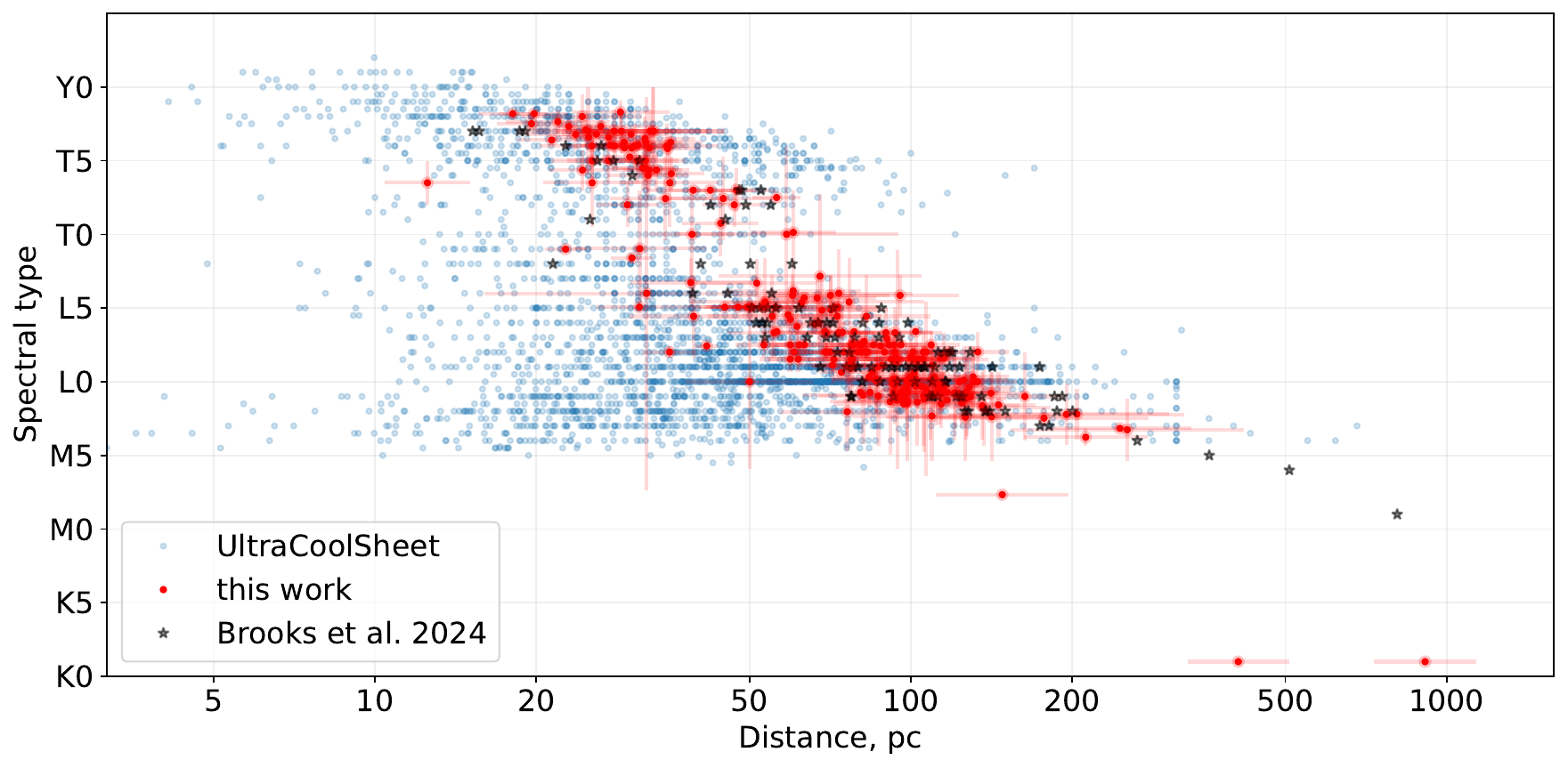}
\includegraphics[width=\columnwidth]{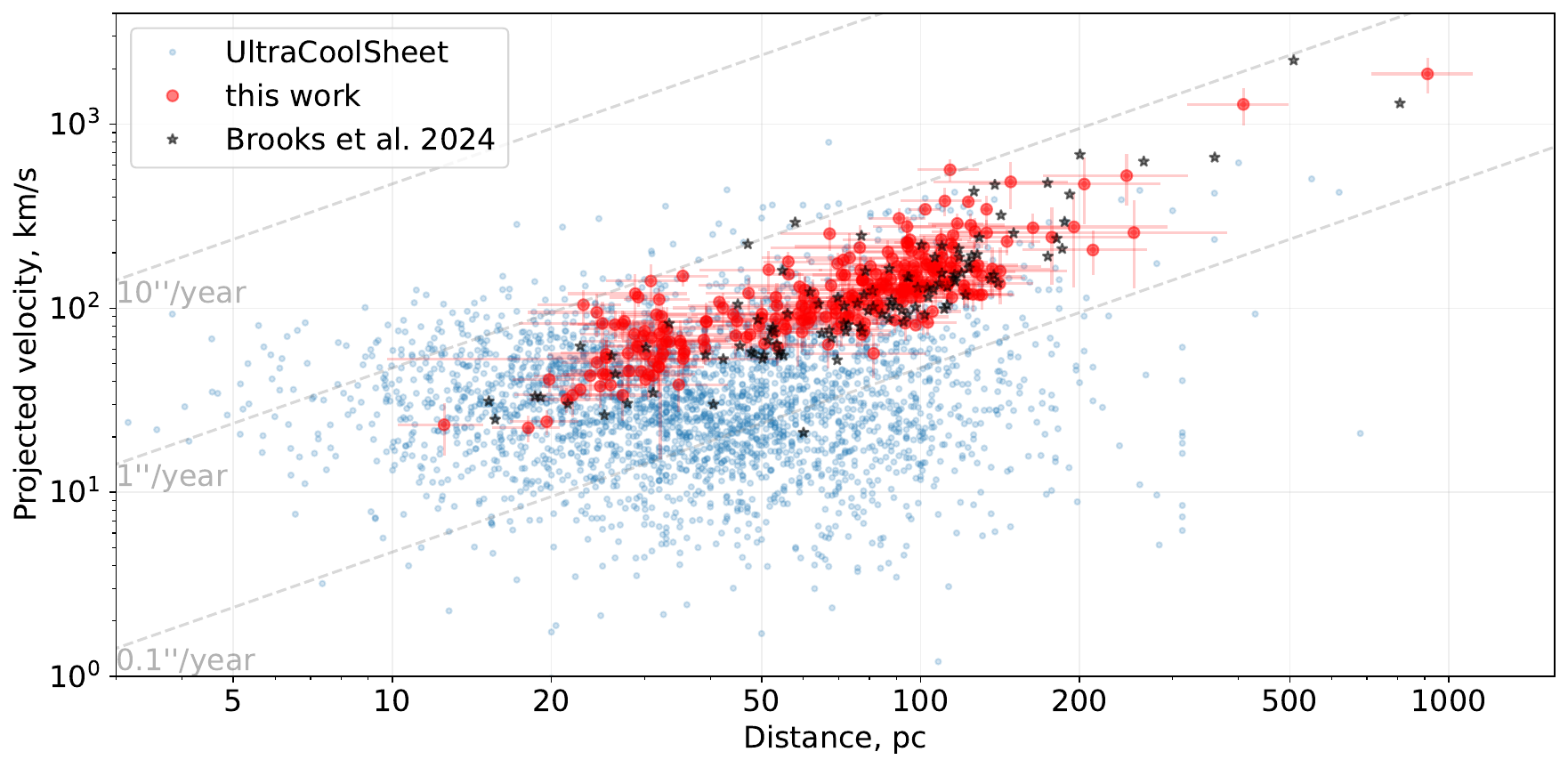}
\caption{\new{Distances and spectral types (left panel), and distances and projected velocities (right panel)
for new moving objects discovered in this work (red points) in comparison to known ultracool dwarfs from UltracoolSheet \citep{ucs} (blue points) and all 118 objects discovered by \citet{brooks_2024} (black stars). The distances for new objects are estimated by interpolation of $M_W2$ absolute magnitudes using nearby stars from 20-pc census of \citet{kirk2024}, and then used for converting proper motions to projected velocities.
Gray diagonal lines in the right panel correspond to fixed values of proper motion.
}
}
\label{fig:spt_dist_vel}
\end{figure*}

\new{
We successfully detected 21,885 moving objects in Section~\ref{sec:algorithm}, including the largest proper motion brown dwarf, WISEA~J085510.74-071442.5\citep{fastest_dwarf}, as well as the fastest known star, Barnard's Star, and the majority of known high proper motion ultracool dwarfs from UltracoolSheet \citep{ucs,ucs1,ucs2}. However, we did not detect anything moving faster than that, despite the algorithm specifically being tailored to be sensitive to even the largest possible proper motions (see Figure~\ref{fig:efficiency}).
}

\new{
Our algorithm facilitated the identification of 21,885 confirmed moving objects from 59,322 candidates through visual inspection. This approach demonstrated a notable advancement over prior methods. For instance, the image-based neural network \texttt{SMDET}, applied to unWISE coadds, identified 1,730 real proper motion objects from 11,900 inspected candidates \citep{brooks_2024}. Additionally, our algorithm uncovered 323 moving objects without SIMBAD and Gaia associations -- nearly three times more than the 118 reported in a similar range of proper motions by \citet{brooks_2024} -- while successfully recovering 38 objects from their sample.
}


\new{
The process of visually inspecting initial candidates and identifying moving objects based on their visual signatures (see Figure~\ref{fig:diff}) closely resembles the approach used by the \textit{Backyard Worlds: Planet 9} citizen science project \citep{kuchner_2017}, which involves tens of thousands of participants and has been ongoing for seven years. However, by leveraging a motion detection algorithm to generate track candidates, applying a machine learning-based artifact rejection routine, and utilizing a compact, Jupyter-based dashboard, our visual inspection process is significantly more fast and efficient.
}


\new{Backyard Worlds: Cool Neighbors \citep{byw_coolneighbors} citizen science project refines the original by also using an input list of positions -- in their case, selected from CatWISE2020\citep{catwise2020} catalogue using machine learning classifier tailored specifically for finding new T and L brown dwarfs, with the expected efficiency of about 0.5\% due to large amount of spurious high proper motion entries polluting the catalogue. In contrast to them, our method is not restricted to finding just brown dwarfs, and is better suited for detecting high proper motion objects which is difficult for CatWISE2020\citep{catwise2020}. Moreover, among 258 new moving objects we detected three that are lacking CatWISE2020  associations, proving that our approach is also complementary to the ones based on that catalogue like \citet{byw_coolneighbors} or \citet{byw_ml_catwise}.
}

\new{
Figure~\ref{fig:galactic} shows sky positions of the newly detected objects in Galactic coordinates. It is consistent with isotropic distribution expected for nearby Galactic objects.
We may qualitatively assess the distances to them by regressing the absolute $W2$ magnitudes of objects with known parallaxes from 20-pc census of stars and brown dwarfs \citep{kirk2024} versus their spectral types, and then applying the same dependency (along with corresponding scatter) for best spectral type estimates from our sample, as described in Section~\ref{sec:new}. It then allows (if we ignore the reddening which is reasonable for such nearby objects) estimating the distance (``photometric distance'') using observed $W2$ magnitudes with formal accuracy typically better than 20\%, as well as estimating projected velocities with comparable accuracy, as shown in Figure~\ref{fig:spt_dist_vel}. Comparison of these photometric distances with astrometric ones for the objects we detected in Section~\ref{sec:new} that lack SIMBAD associations but do have Gaia DR3 or DR2 counterpart is shown in left panel of Figure~\ref{fig:dist_gaia}. While Gaia detected objects all have spectral types earlier than M9, we may still see that photometric distances are, within the errors, compatible with the parallaxes, for most of the colder objects. Starting with approximately M3 spectral type and earlier, the estimation becomes biased and significantly (by about factor of two) overestimates the distance for late K types.
We cannot easily guess the reason for this behaviour, but it may explain the anomalously high distances and projected velocities for two hottest objects in our actual new objects sample shown in Figure~\ref{fig:spt_dist_vel}.
The rest of the objects there should not be affected by this problem, and we may consider their photometric distances reliable.
On the other hand, the agreement between the effective temperatures estimated in Section~\ref{sec:new} with the ones derived using Gaia DR3 GSP-Phot pipeline \citep{gaia_gspphot} is sufficiently good up to at least T$_{\rm eff}=4300$K, or K6 spectral type (see right panel of Figure~\ref{fig:dist_gaia}.
}

\new{
Among 258 new objects we detected all except 6 are compatible with being ultracool dwarfs colder than M7 type.
At least 62 are consistent with being brown dwarfs of types later than T0, i.e. having $T_{\rm eff}<1300$K or $W1-W2>0.7$. Among them, 33 have effective temperatures estimated using BT-Settl spectral grid, and thus may be considered most reliable new T brown dwarf candidates that require further spectroscopic study for their confirmation. They all have estimated distances of 20--40 pc, with the closest one -- CWISE~J154802.28+565802.2 -- being at a $18.1\pm2.6$pc distance and with T8 estimated spectral type.
We did not detect any new hypervelocity star \citep{hypervelocity} (two objects with formal projected velocity exceeding 1000 km/s in Figure~\ref{fig:spt_dist_vel} are most probably due to overestimated distance, see above), and no brown dwarf candidate with projected velocity significantly greater than $\sim500$ km/s.
}

\new{
Relatively small proper motions of new objects that all move slower than 1$''$/year argue against Solar System origin for any of them. Indeed, even a hypothetical brown dwarf companion of the Sun in in the vicinity of the aphelion of a very eccentric orbit, would exhibit a proper motion of about 4.6$''$/year \citep{malkov_sun_companion}. Bodies of the inner Solar System all have even faster motions, again incompatible with the ones of newly found objects.
}

\new{
In our dataset we detected four of the spectroscopically confirmed T subdwarfs recently reported by \citet{burg2024}, for two of them we have VOSA-estimated effective temperatures that agree well with published values. Figure~\ref{fig:colors} shows their positions in the two-color diagrams, as well as in the reduced $J$ magnitude versus $J-W2$ color one, alongside with the rest of our objects. A number of them satisfy the criteria for selection of T subdwarf candidates that \citet{burg2024} used, so we may assume that at least part of them might also be subdwarfs. As shown in Figure~\ref{fig:spt_comp} we may expect our spectral types to be underestimated for them, but as Figure~\ref{fig:dist_gaia} demonstrates their effective temperatures should be still estimated correctly. The accuracy and spectral coverage of the photometric data we assembled are not enough to reliably identify subdwarfs through their sub-solar metallicity using model fitting, so we did not perform any additional study for these potential subdwarf candidates.
}

\new{
As Figure~\ref{fig:spt_dist_vel} shows, the new objects we detected are on the faster / colder / more distant side of the overall distribution of already known ultracool dwarfs, while not being the extreme ones due to limitations of unTimely catalogue. Their overall spread is very similar  to the sample of 118 ones discovered by \citet{brooks_2024}, and is defined by the selection effects -- minimal detectable proper motion from one side, and the fact that most of faster-moving objects were already detected in AllWISE \citep{2014ApJ...783..122K} on other side -- the same for both this work and \citet{brooks_2024}.
On the other hand, our results suggest that there is still a place for new discoveries for brown dwarfs within this range of parameters, even before new instrumentation like James Webb Space Telescope, Vera C. Rubin Observatory or Nancy Grace Roman Space Telescope will enable new horizons in brown dwarf science \citep{lsst_bd_parallaxes, 2024AJ....168..179L, bd_populations}.
}



\section{Conclusions}
\label{sec:conclusions}

We performed a systematic search for the objects with high proper motions in $W2$ band WISE data using recently released unTimely epochal catalogue, and successfully identified \new{21,885} objects moving faster than \new{0.3}$''$/year. Our estimations of their proper motions are in a good agreement with Gaia DR3 measurements where they are available. We did not detect any object moving faster than approx. 10$''$/year, but were able to successfully recover both WISEA~J085510.74-071442.5 \new{brown dwarf with largest known proper motion}
and most of other known rapidly moving normal stars and nearby ultracool dwarfs. We also identified \new{258} objects that do not have associated records in SIMBAD database or Gaia catalogues, and had not been published previously. We assembled multi-wavelength information for them by cross-matching their trajectories with large optical and infrared catalogues, and used their colors to estimate their \new{effective temperatures,} spectral types, distances and projected velocities. \new{252 are compatible with being ultracool dwarfs, and at least 33 of them are reliable T brown dwarf candidates with estimated distances closer than 40 pc and effective temperatures colder than 1300 K.}


\begin{acknowledgements}
This work was co-funded by the EU and supported by the Czech Ministry of Education, Youth and Sports through the project CZ.02.01.01/00/22\_008/0004596.
This work has benefitted from The UltracoolSheet at \url{http://bit.ly/UltracoolSheet}, maintained by Will Best, Trent Dupuy, Michael Liu, Aniket Sanghi, Rob Siverd, and Zhoujian Zhang, and developed from compilations by Dupuy \& Liu (2012, ApJS, 201, 19), Dupuy \& Kraus (2013, Science, 341, 1492), Deacon et al. (2014, ApJ, 792, 119), Liu et al. (2016, ApJ, 833, 96), Best et al. (2018, ApJS, 234, 1), Best et al. (2021, AJ, 161, 42), Sanghi et al. (2023, ApJ, 959, 63), and Schneider et al. (2023, AJ, 166, 103).

This publication makes use of VOSA, developed under the Spanish Virtual Observatory (https://svo.cab.inta-csic.es) project funded by MCIN/AEI/10.13039/501100011033/ through grant PID2020-112949GB-I00.
VOSA has been partially updated by using funding from the European Union's Horizon 2020 Research and Innovation Programme, under Grant Agreement No 776403 (EXOPLANETS-A)
\end{acknowledgements}

\bibliographystyle{aa} 
\bibliography{main} 

\end{document}